\documentclass[english,11 pt]{article}
\usepackage[T1]{fontenc} 
\usepackage[latin9]{inputenc}
\usepackage{graphicx}
\usepackage{babel}
\usepackage{amsmath}
\usepackage{multicol}
\makeatletter

\newcommand{\fig}[1]{Figure (\ref{#1})}
\newcommand{\eq}[1]{Eq. (\ref{#1})}
\newcommand{\eqs}[1]{Eqs. (\ref{#1})}
\newcommand{\se}[1]{Sec. (\ref{#1})}
\newcommand{\ch}[1]{Chapter (\ref{#1})}
\newcommand{\la}[1]{ \label{#1}}

\renewcommand{\b}{\beta}
\newcommand{\e}{\epsilon}
\renewcommand{\r}{\mathbf{r}}

\renewcommand{\k}{\mathbf{k}}

\newcommand{\bsubs}{\begin{subequations}}
\newcommand{\esubs}{\end{subequations}}

\newcommand{\be}{\begin{equation}}
\newcommand{\ee}{\end{equation}}

\newcommand{\bea}{\begin{eqnarray}}
\newcommand{\eea}{\end{eqnarray}}

\begin{document}

\title{Theories of Matter: Infinities and Renormalization\\
}
\author{ Leo P. Kadanoff\\
The James Franck Institute\\
The University of Chicago\\
Chicago, Illinois, USA\\
and\\
The Perimeter Institute\\
Waterloo, Ontario, Canada\\
email: leop@UChicago.edu\\
Draft Version2.0\footnote{Copies of this paper have been published on arXiv and on the author's web site:  jfi.uchicago.edu$\backslash \sim$ leop$\backslash$. }}

\maketitle
\begin{abstract}

This paper looks at the theory underlying the science of materials from the perspectives of physics, the history of science, and the philosophy of science.  We are particularly concerned with the development of understanding of the thermodynamic phases of matter.  

The question is how can matter, ordinary matter, support a diversity of forms.   We see this diversity each time we observe ice in contact with liquid water or see water vapor (steam) rise from a pot of heated water. Different phases can be qualitatively different. For example, walking on ice is well within human capacity, but walking on liquid water is proverbially forbidden to ordinary humans.  These differences have been apparent to humankind for millennia, but only brought within the domain of  scientific understanding since the 1880s.   
 
The nature of the phases is brought into the sharpest focus in phase transitions:  abrupt changes from one phase to another and hence changes from one behavior to another.  A first-order phase transition involves a discontinuous jump in some statistical variable.  The property with the jump is called the {\em order parameter}. Each phase transition has its own order parameter.  The possible order parameters range over a tremendous variety of physical properties.  These properties include the density of a liquid-gas transition, the magnetization in a ferromagnet, the size of a connected cluster in a percolation transition, and a condensate wave function in a superfluid or superconductor. A continuous transition occurs when the discontinuity in the jump approaches zero.   This article starts with the development of mean field theory as a basis for a partial understanding of phase transition phenomena.    It then goes on to the limitations of mean field theory and the development of very different supplementary understanding through the {\em renormalization group} concept.

 Throughout, the behavior at the phase transition is illuminated by an ``extended singularity theorem'', which says that a sharp phase transition only occurs in the presence of some sort of infinity in the statistical system.  The usual infinity is in the system size.  Apparently this result caused some confusion at a 1937 meeting celebrating van der Waals, since mean field theory does not respect this theorem.  In contrast, renormalization theories can make use of the theorem.  This possibility, in fact, accounts for some of the strengths of renormalization methods in dealing with phase transitions. The paper outlines the different ways phase transition phenomena reflect the effects of this theorem.

\end{abstract}
\tableofcontents
\newpage{ }
\section{Introduction}
\subsection{The discovery and invention of materials}
 From the ``stone age'' through the present day humankind has made use of the materials available to us in the earth and equally to materials we could manufacture for further use. Jared Diamond\cite{Diamond}, for example, in his book {\em Guns, Germs and Steel,} has pointed out how crucial metalworking was to the spread of European power. 

However, it is only in relatively recent times that we were able to bring scientific understanding of the inner workings of materials to human benefit. The latter half of the nineteenth century brought the beginning of two major theoretical advances to the science of materials, advances that would deepen and grow into the twentieth century so that today we can boast of a fundamental understanding of the main properties of many materials.    These advances are: a theory of statistical physics developed initially by Rudolf Clausius\cite{BrushKinetic}, J. C. Maxwell\cite{maxwell-Gases}, and Ludwig Boltzmann\cite{BrushKinetic,SBrush} and an understanding of the different phases of matter based in part upon an understanding of the changes from one phase\footnote{The word ``phase'' is interesting.  According to the{\em \/ Oxford Dictionary of Word Histories} (and the {\em Oxford English Dictionary}) it entered English language in the nineteenth century to describes the phases of the moon. The {\em Oxford English Dictionary~}lists a very early use in J.~Willard Gibbs' writings about thermodynamics as the ``phases of matter''.   Apparently Gibbs then extended the meaning to get ``extension in phase'' that then got further extended into the modern usages ``phase transition'' and ``phase space''.} to another.  These changes are called phase transitions and our understanding of them is based upon the work of Thomas Andrews\cite{Andrews}, Johannes van der Waals\cite{Levelt-Sengers} and Maxwell.   It is the purpose of this paper to develop a description of the development of these basic ideas from the 1870s to the last quarter of the twentieth century. 

Much of our first principles understanding of materials is based upon the fact that they are composed of many, many atoms, electrons, and molecules.  This means  that we cannot hope or wish to follow any motion of their individual constituents, but that instead we must describe their average or typical properties through some sort of statistical treatment.  Further, though we should believe that the properties of these materials is built upon the properties of the constituents, we should also recognize that the properties of a huge number of constituents, all working together, might be quite different from the behaviors we might infer from thinking about only a few of these particles at once.  P. W. Anderson has emphasized the difference by using the phrase ``More is Different''\cite{PWA,reviewPWA}.  

There are two surprising differences that have dominated the study of materials: irreversibility and the existence of sharply distinct thermodynamic phases.   First, irreversibility:   Even though the basic laws of both classical mechanics and quantum mechanics are unchanged under a time reversal transformation, any appropriately statistical treatment of systems containing many degrees of freedom will not show such an invariance.  Instead, such systems tend to flow irreversibly toward an apparently unchanging state called {\em statistical equilibrium}.  This flow was recognized by Clausius in his definition of entropy,  detailed by Boltzmann in his gas dynamic definition of the Boltzmann equation, and used by  Gibbs\cite{MR} in his definitions of thermodynamics and statistical mechanics.   It is a behavior that is best recognized as a property of a limiting case, either of having the number of degrees of freedom become infinite or having an infinite observation time.  Time reversal asymmetry is not displayed by any finite system described over any finite period of time\footnote{Particle physics does show a very weak time reversal asymmetry discovered by James W. Cronin and Val L. Fitch\cite{CF}, but this asymmetry is immaterial for all mundane phenomena.}.

This paper is concerned with thermodynamic equilibrium resulting from this irreversible flow. It focuses upon another property of matter that, as we shall see, only fully emerges when we consider the limiting behavior as an equilibrium material becomes infinitely large.     This property is the propensity of matter to arrange itself into different kinds of structures that are amazingly diverse and beautiful. These structures are called {\em thermodynamic phases}. \fig{waterstructure} illustrates three of the many thermodynamic phases formed by water.   Water has many different solid phases.Other fluids form liquid crystals, in which we can see  macroscopic manifestations of the shapes of the molecules forming the crystals. The alignment of atomic spins or electronic orbits can produce diverse magnetic materials, including ferromagnets, with their substantial magnetic fields, and also many other more subtle forms of magnetic ordering. Our economic infrastructure is, in large measure,  based upon the various phase-dependent capabilities of materials to carry electrical currents:  from the refusal of insulators, to the flexibility of semiconductors,  to the substantial carrying capacity of conductors, to the weird resistance-free behavior of superconductors. This flow, and other strange properties of superconductors, are manifestations of the subtle behavior of quantum systems, usually only seen in microscopic properties, but here manifested by these materials on the everyday scales of centimeters and inches. I could go on and on.  The point is that humankind has, in part, understood these different manifestations of matter, manifestations that go under the name ``thermodynamic phases''. Scientific work has produced at least a partial understanding of how the different phases change into one another.    This article is a brief description of the ideas contained in the science of such things.

As is the case with irreversibility, the differences among solid, liquid, and gas; the distinctions among magnetic materials and between them and  non-magnetic materials; and the differences  between normal materials and superfluids are all best understood as distinctions that apply in the limit in which the number of molecules is infinite.  For any finite body, these distinctions are blurry with the different cases merging into one another. Only in the infinite limit can the sharp distinction be maintained.  Of course, our usual samples of everyday materials contain a huge number of molecules, so the blur in the distinction between different phases is most often too fine for us to discern.  However, if one is to set up a theory of these materials, it is helpful to respect the difference between finite and infinite. 
       
\subsection{Different phases; different properties} 
Three of the phases of water are illustrated in \fig{waterstructure}, which depicts a snowflake (a crystal of a solid phase of water) and a splashing of the liquid phase.  A third phase of water, the low density phase familiar as water vapor or steam, exists  in the empty-looking region above and around the splashing liquid.  

Different phases of matter have qualitatively different properties.    As you see, ice forms beautiful crystal structures. So do other solids,  each with its own  characteristic shape and form.  Each crystal picks out particular spatial directions for its crystal axes.  That selection occurs because of forces produced by the interactions of the microscopic constituents of the crystals.  Crystalline materials are formed at relatively low temperatures.  At such temperatures,  microscopic forces tend to line up neighboring molecules and  thereby produce strong correlations between the orientations of close neighbors.  Such correlations extend through the entire material, with each molecule being lined up by several neighbors surrounding it, thereby producing an ordering in the orientation of the molecules that can extend over a distance a billion times larger than the distance between neighboring atoms.  This orientational order thus becomes visible in the macroscopic structure of the crystal, as in \fig{waterstructure}, forming a macroscopic manifestation of the effects of microscopic interactions.   

The same materials behave differently at higher temperature.  Many melt and form a liquid.  The long-range orientational order disappears.  The material gains the ability to flow. It loses its special directions and gains the full rotational symmetry of ordinary space.   

Many of the macroscopic manifestations of matter can be characterized as having {\em broken symmetries.}  Phases other than simple vapors or liquids break one or more of the characteristic symmetry properties obeyed by the microscopic interactions of the constituents forming these phases.    Thus, a snowflake's outline changes as it is rotated  despite the fact that the molecules forming it can, when isolated, freely rotate.

 \begin{figure}
\hskip  30 pt
\includegraphics[height=4cm ]{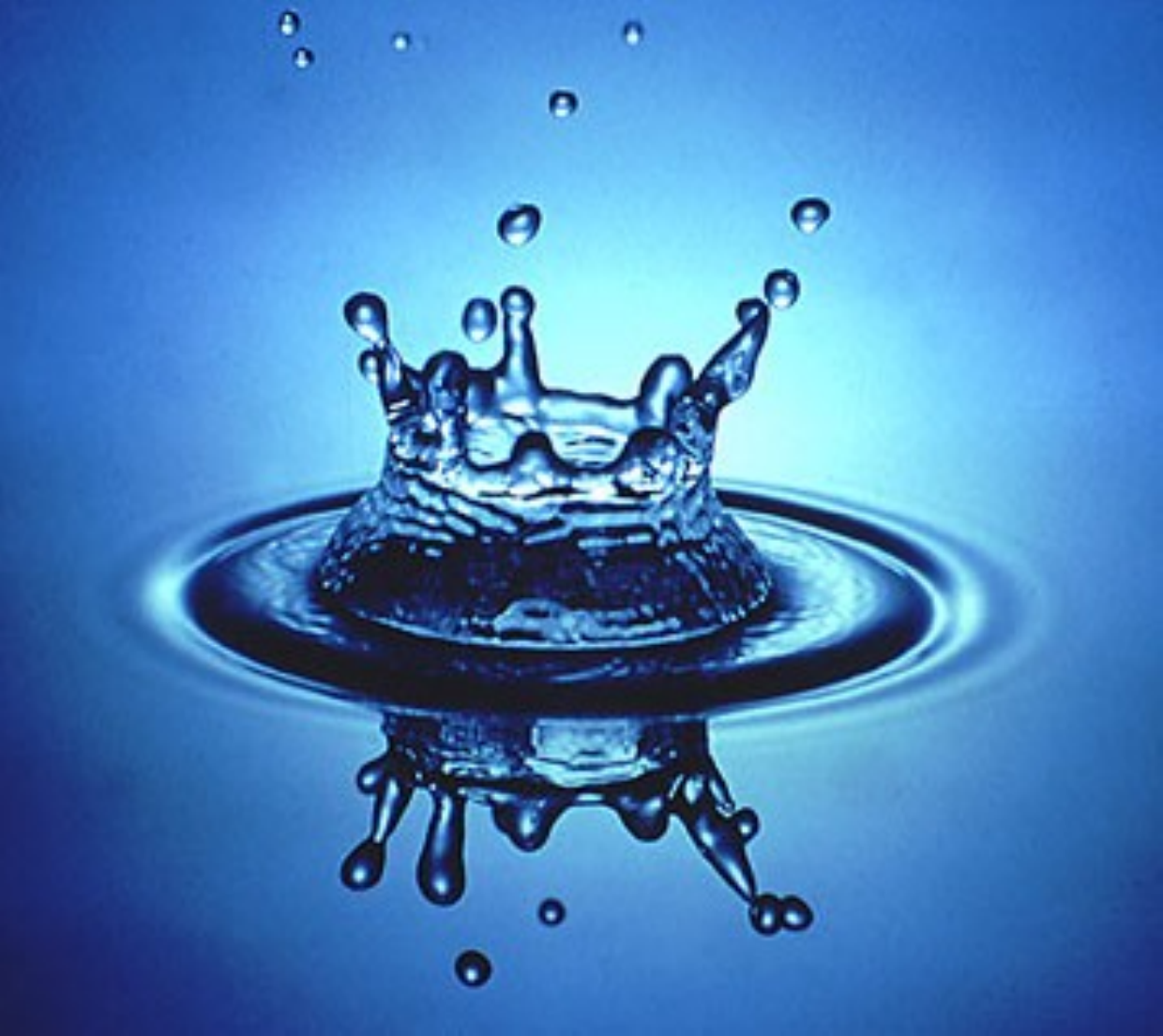}
\hskip  40 pt
\includegraphics[height=4cm ]{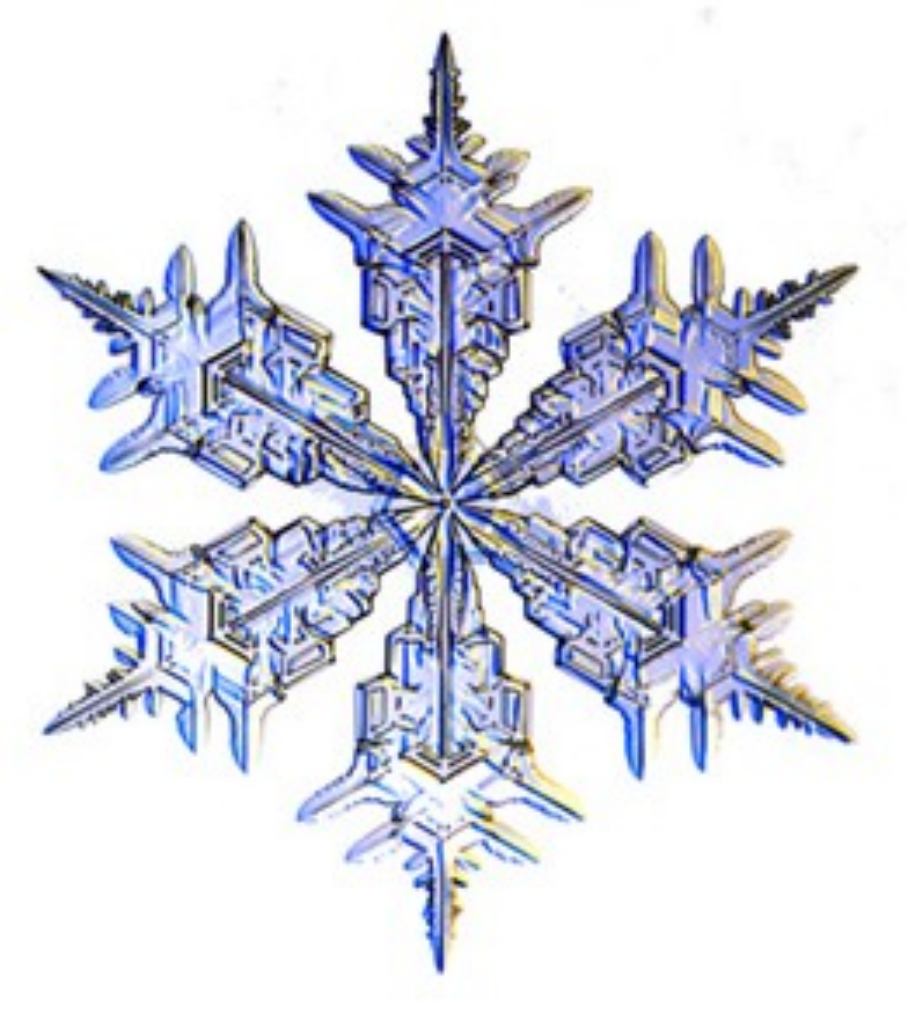}
\caption{Splash and snowflake.  This picture is intended to illustrate the qualitative differences between the fluid and solid phases of water. On the left is liquid water, splashing up against its vapor phase.  Its fluidity is evident.  On the right is a crystal of ice in the form of a snowflake.  Note the delicate but rigid  structure, with its symmetry under the particular rotations that are multiples of sixty degrees.  }  
\la{waterstructure}
\end{figure}

The same materials behave differently at higher temperature.  Many melt and form a liquid.  The long-range orientational order disappears.  The material gains the ability to flow. It loses its special directions and gains the full rotational symmetry of ordinary space.   

Many of the macroscopic manifestations of matter can be characterized as having {\em broken symmetries.}  Phases other than simple vapors or liquids break one or more of the characteristic symmetry properties obeyed by the microscopic interactions of the constituents forming these phases.    Thus, a snowflake's outline changes as it is rotated  despite the fact that the molecules forming it can, when isolated, freely rotate.

\subsubsection{Broken symmetries and order parameters.} 
A previous publication\cite{I} describes the development of the theory of phase transitions up to and including the year 1937.   In that year, Lev Landau\cite{Landau} put together a theoretical framework that generalized previously existing mean field theories of phase transitions.  In Landau's approach, a phase transition manifests itself in the breaking of a mathematical symmetry.  This breaking is, in turn, reflected in the behavior of an {\em order parameter} describing both the magnitude and nature of the broken symmetry.  Two different phases placed in contact are seen to be distinguished by having different values of the order parameter.   For example, in a ferromagnet the order parameter is the vector magnetization of the material.   Since the spins generate magnetic fields, this alignment is seen as a large time-independent magnetic field vector, pointing in some particular direction.   The order parameter in this example is then the vector describing the orientation and size of the material's magnetization and its resulting magnetic field.  Note that the {\em orientation} of the order parameter describes the way in which the symmetry is broken, while the {\em magnitude} of this parameter describes how large the symmetry breaking might be.   

Other order parameters describe other situations. A familiar order parameter  characterizes the difference between the liquid and the vapor phases of water.  This parameter is simply the mass density, the mass per unit volume, minus the value of that density at the critical point.  This parameter takes on positive values in the liquid phase and negative ones in the vapor phase.  These phases are clearly exhibited in processes of condensation, or boiling, or when the two phases are placed in contact with one another.    In superfluid examples, ones in which there is particle flow without dissipation,  a finite fraction of the particles are described by a single, complex wave function.   The order parameter is that wave function. 

The alignment of atomic spins or electronic orbits can produce diverse magnetic materials, including ferromagnets, with their substantial magnetic fields, and also many subtler forms of magnetic ordering. A crystal in which the magnetization can point in any direction in the three-dimensional space conventionally labeled by the X, Y, and Z axes  is termed an ``XYZ'' ferromagnet.   Another kind of ferromagnet is called an ``XY'' system, and is one in which internal forces within the ferromagnet permit the magnetization to point in any direction in a plane.  The next logical possibility is one in which there a few possible axes of magnetization, and the magnetization can point either parallel or antiparallel to one of these axes.    A simple model describing this situation is termed an Ising model, named after the physicist Ernst Ising\cite{Ising}, who studied it in conjunction with his adviser Wilhelm Lenz\cite{Lenz,Niss}(See \ch{IsingM}).

\subsubsection{Dynamics and equilibrium}
We have already mentioned the propensity of condensed matter systems to approach  an unchanging state called thermodynamic equilibrium.  This approach can be either extremely fast or extremely slow, depending upon the situation.  When an electrical current in a metal is set in motion by an applied voltage it can take as little as a millionth of a billionth of a second for the current to reach close to its full value.  On the other hand, impurities may take years to diffuse through the entire volume of a metal.  Because of this very broad range of time scales, and because of the wide variety of mechanisms for time-dependence, dynamics is a very complex subject. 

Equilibrium is much simpler. The equilibrium state of a simple material  is characterized by a very few environmental parameters.   These parameters correspond to a knowledge of the amount or availability of the conserved (i.e., time-independent) quantities describing the system.  To describe a container filled with water, one needs to know its temperature, the volume of the container, and the applied pressure. One can then directly  calculate the the mass of the water in the volume using the data from what is called the {\em equation of state}. 

 By the time our story begins, in the early twenthieth century, {\em kinetic theory} will be well-developed as a description of common  gases and liquids.  That theory includes the statement that molecules in these fluids are in rapid motion, with a kinetic energy proportional to the temperature.   {\em Thermodynamics} provides a more broadly applicable theory,   brought close to its present state by J. Willard Gibbs in 1878\cite[p. 55-371]{Gibbs-CW,MR}.   That discipline provides relations among the properties of many-particle systems based upon conservation of energy and the requirement that the system always develops in the direction of thermodynamic equilibrium.  Thermodynamic calculations are based upon thermodynamic functions, for example the Helmholtz free energy used in this paper. This function depends upon the material's temperature, volume, and the number of particle of various types within it  One can calculate all kinds of other properties by calculating derivatives of the free energy with respect to its variables.  The free energy has the further important property that the actual equilibrium configuration of the system will produce the minimum possible value of the free energy.  Any other configuration at that temperature, in that volume, with those constituents will necessarily have a higher free energy. 
 
To understand phase transitions, one has to go beyond thermodynamics,  which concerns itself with relations among macroscopic properties of materials, to the subject of {\em statistical mechanics} that defines the probabilities for observing various phenomena in a material in thermodynamic equilibrium, starting from a microscopic description of the behavior of the materials' constituents. Statistical mechanics further differs from thermodynamics in that the latter treats only summed or gross properties of matter, while the former looks to the individual constituents and asks about the relative probabilities for their different configurations and motions.     
 
 To use statistical mechanics, in principle, all one needs to know is a function called the {\em Hamiltonian}, which gives the system's energy as a function of the coordinates and momenta of the particles.\footnote{This function is named after William Rowan Hamilton who described how to formulate classical mechanics using this Hamiltonian function.} In practice, for many-particle systems, the actual calculations are sufficiently hard so that in large measure they only became possible after World War II.  

Statistical mechanics is classically defined by using a {\em phase space} given by the momenta and coordinates of the particles in the system.  The state of a particular classical system is then defined by giving a point in that space.  The basics of statistical mechanics were put forward by Boltzmann\cite[page 8 and chapter 7]{CC},\cite{Uffink},\cite{GG} and then clearly stated by Gibbs\cite{GibbsE} in essentially the same form as it is used today. It describes the probability for finding a classical system in a particular configuration, $c$, in the phase space.  The probability of finding the system in a small volume of phase space, $d\Omega$, around configuration $c$ is given by       
\be
e^{-({\mathcal{H}(c)-F})/{T}}~ d\Omega
\la{MB}
\ee       
Here, $\mathcal{H}(c)$ is the energy in this configuration, $T$ is the absolute temperature in energy units, and $F$ is a constant, which turns out to be the free energy.  Since the total probability of all configurations of the system must add up to unity, one can determine the free energy by
\be
e^{-F/T}= \int ~ d\Omega~e^{-({\mathcal{H}(c)})/{T}} 
\la{free}
\ee
where the integral covers all of phase space.   

This probability formulation, the basis of all of statistical mechanics, is known as the Maxwell-Boltzmann distribution, to most physicists and chemists,  or the Gibbs measure, to most mathematical scientists. 

This probability of \eq{MB}  describes a collection of identical materials arrayed in different  configurations.  Gibbs calls the ``ensemble'' of configurations thus assembled a ``canonical ensemble\footnote{The word  ``canonical'' seems to be an somewhat old-fashioned  usage for something set to a given order or rule. The {\em Oxford English Dictionary} traces it back to Chaucer.}.''

\subsubsection{Phase transitions}

 \begin{figure}
\begin{multicols}{2}
\includegraphics[height=6cm ]{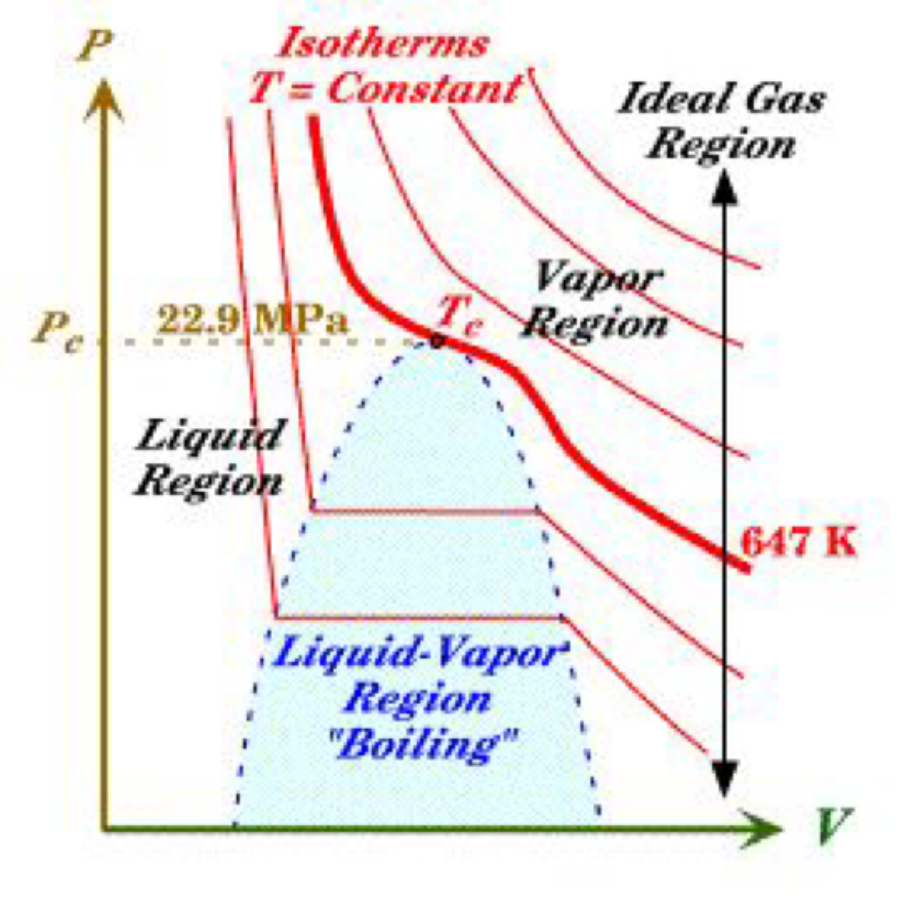}
\vskip 30 pt
\caption{Cartoon PVT diagram for  water.   Each curve describes how the pressure depends upon volume for a fixed temperature.  Note the figures for critical temperature and pressure on this diagram.  They apply to water. The corresponding figures of carbon dioxide are 31.1$^\circ$  C and 73 atmospheres. These values are more easily accessible to experiments than the ones for water.      }  
\la{PVT}
\end{multicols}
\end{figure}

A treatment of phase transitions may properly start with the 1869 experimental studies of Andrews\cite{Andrews}, who investigated the phase diagram of carbon dioxide and thereby discovered the qualitative properties of the liquid-gas phase transition.   His results, as illuminated by the theoretical work of van der Waals and Maxwell\cite{Levelt-Sengers}, are shown in \fig{PVT}.   This plot gives the pressure as it depends upon volume in a container with a fixed number of particles.  Each curve shows the behavior of the pressure at a given value of the temperature.   The two curves at the bottom show results quite familiar  from our experience with water.   At high pressures one has a liquid and the liquid is squeezed into a relatively small volume.  Since the density of the liquid is the (fixed) number of particles divided by a varying volume, this high pressure region is one of high density.   The curve moves downward showing a reduced pressure as the liquid is allowed to expand.   At a sufficiently low pressure, the liquid starts to boil, and thereby further reduce its density until a sufficiently low volume is reached so that it has attained the density of vapor. The boiling is what is called a {\em first-order phase transition}.  The boiling occurs at a constant pressure and has liquid and vapor in contact with one another.  Then after the vapor density is reached, additional expansion produces a further reduced density.     At somewhat higher temperatures this same scenario is followed, on a higher curve,   except that the region of boiling and its connected jump in density is smaller.   Andrews' big discovery was that at a sufficiently high temperature, the jump in density disappears and the fluid goes from high pressure to low without a phase transition.   This disappearance of the first-order phase transition occurs at what is called a {\em critical point}.     The disappearance itself is called a {\em continuous phase transition.}

\subsubsection{The first mean field theory}
In his thesis of (1873), Johannes van der Waals\cite{Waals} put together an approximate theory of the behavior of liquids using arguments based in ideas of the existence of molecules. The very existence of molecules was an  idea  then current, but certainly not proven.  Van der Waals started from the known relation between the pressure and the volume of a perfect gas, i.e., one that has no interactions between the molecules.  Expressed in modern form, the relation is  
\be
p=  T~N /V
\la{perfect}
\ee
Here, $p$ is the pressure, $V$ is the volume of the container, $N$ is the number of molecules within it, $T$ is the temperature expressed in energy units\footnote{I use energy units in order to write fewer symbols.  It is more conventional to write, instead of $T$,  $kT$, where $k$ is the Boltzmann constant.  }. This {\em equation of state} relates the pressure, temperature, and density of a gas in the dilute-gas region in which we may presume that interactions among the atoms are quite unimportant.   It says that the pressure is proportional to the temperature,  $T$, and to the density of particles, $N/V$.            This result is inferred by ascribing an average kinetic energy to each molecule proportional to $T$ and then calculating the transfer of momentum per unit area to the walls.  The pressure is this transfer per unit time.  Of course,  \eq{perfect} does not allow for any phase transitions.

Two corrections to this law were introduced by van der Waals to  estimate how the interactions among the molecules would affect the properties of the fluid.  

First, he argued that the molecules could not approach each other too closely because of an inferred short-ranged repulsive interaction among the molecules. This effect should reduce the volume available to the molecules by an amount proportional to the number of molecules in the system.  For this reason,  he replaced $V$ in \eq{perfect} by the available volume, $V-Nb$, where $b$ would be the excluded volume around each molecule of the gas.  

The second effect is more subtle.    The pressure, $p$, is a force per unit area produced by the molecules hitting the walls of the container.  However, van der Waals inferred that there was an attractive interaction pulling each molecule towards its neighbors. This attraction is the fundamental reason why a drop of liquid can hold together and form an almost spherical shape.   As a molecules move toward a wall, it is pulled back and slowed by the molecules left behind. Because of this reduced speed, it imparts less momentum to the walls than it would otherwise.    The equation of state contains the pressure as measured at the wall, $p$.  This pressure is the one produced inside  the liquid, $NT/(V-Nb)$, minus  the correction term coming from the interaction between the molecules near the walls.   That correction term is proportional to  the density of molecules squared. In symbols  Van der Waals'  corrected expression for the pressure is thus
\be
p= NkT/(V-Nb)-a(N/V)^2
\la{Waals}
\ee
Here, $a$ and $b$ are parameters that are different for different fluids and $N/V$ is the density of molecules. 

\eq{Waals} is the widely used van der Waals equation of state for a fluid. Because it takes into account average forces among particles, we describe it and similar equations as the result of a {\em mean field theory}.  This equation of state can be used to calculate the particle density, $\rho=N/V$, as a function of temperature and pressure.  It is a cubic equation for $\rho$ and thus  has at most  three solutions.

 \begin{figure}
\begin{multicols}{2}
\includegraphics[height=6cm ]{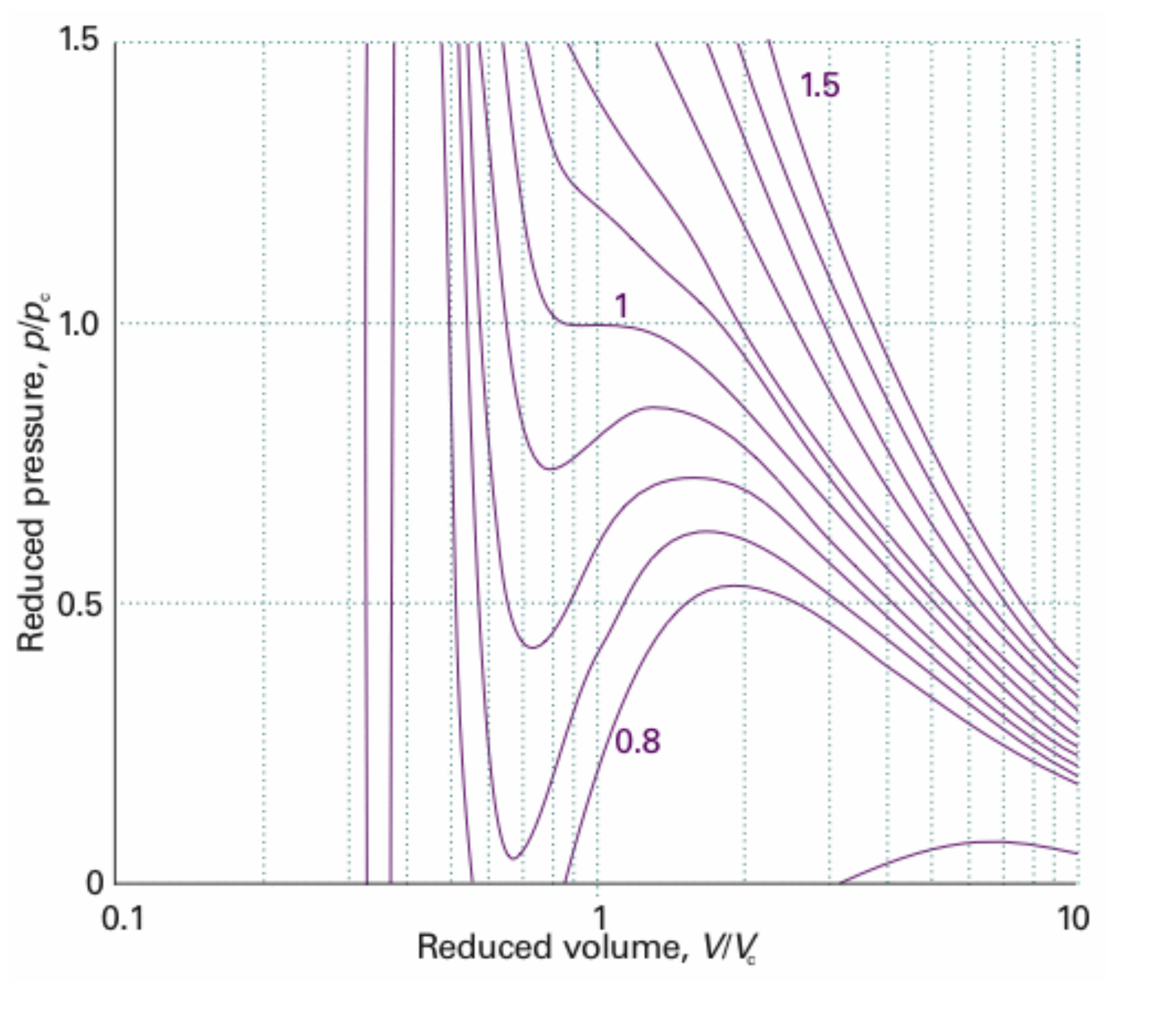}
\caption{PVT curves predicted by the theory set up by van der Waals.  The fluid is mechanically unstable whenever the pressure increases as the volume increases.  }  
\la{PVTv}
\end{multicols}
\end{figure}

\subsubsection{Maxwell's improvement}
The equation of state proposed by van der Waals is plotted in \fig{PVTv}. Each curved line shows the dependence of pressure on volume.  This equation of state has a major defect: it shows no boiling region.   Worse yet, it contains regions in which, at fixed temperature and numbers of particles,  the pressure increases as the volume increases.  This situation is unstable.  If the fluid finds itself in a region with this kind of behavior,  the forces within it will cause it to separate into two regions, one at a high density the other at a lower one.   In fact, exactly this kind of separation does happen in the boiling process in which a lower density vapor is in contact with a high density liquid.

The instability just described is termed a mechanical instability.  It can be triggered by a fluctuation in which two nearby pieces of fluid acquire a density slightly different from that of the fluid elements around them.  J. C. Maxwell\cite{Maxwell, Maxwell1} in 1874 and 1875 recognized this instability and also a somewhat larger region of instability triggered by processes in which nearby pieces of the acquired got a density appropriate for a  thermodynamic equilibrium phase of the fluid, one region having a vapor density, the other a liquid density.   Maxwell identified this larger region of instability with boiling, and drew a phase diagram like that in \fig{PVT}.   Note that this figure has a completely flat portion of the constant temperature lines to represent the predicted boiling of the Maxwell's theory.  We shall hear more of this Maxwell construction in \ch{1937}. 

Maxwell's result gives a qualitative picture of the jump in density between the two phases over a quite wide of temperatures.  For the purposes of this paper, however, the most important region is the one near the critical point in which the jump is small.  According to the theory, as the jump in density, $\rho=N/V$, goes to zero, it shows a behavior    
\be
\rho_{liquid}-\rho_{gas} = \text{ constant}~\times~ (T_c-T)^\b  
\la{beta}
\ee 
where $T_c$ is the critical temperature and $\b$ has the value one half.  Andrews' data does fit a form like this, however with a value of the exponent, $\b$, much closer to one third than one half.  Later on, this discrepancy will become quite important.  

Despite the known discrepancy between mean field theory and experiment in the region of the critical point, few scientists focused upon this issue in the years in which mean field theory was first being developed.   There was no theory or model that yielded \eq{beta} with any power different from one half,  so there was no focus for anyone's discontent.  Thomas Kuhn\cite{Kuhn} has argued that an old point of view will continues on despite evidence to the contrary if there is no replacement theory.

Following soon after van der Waals, many other scientists developed mean field theories, applying them to many different kinds of phase transitions.  All these theories have an essential similarity.  They focus upon some property of the many-particle system that breaks some sort of global symmetry\footnote{In fact, the liquid-gas case is one of the most subtle of the phase transitions since the symmetry between the two phases, gas and liquid, is only an approximate one. In magnets and most other cases the symmetry is essentially exact, before is it broken by the phase transition.}.   In mean field calculations, the ordering in one part of the system induces ordering in neighboring regions until ordering is spread through the entire system.  Thus mean field theory calculations are always descriptive of symmetry breaking and the induced correlations that carry the symmetry breaking through the material.  These calculations are then most relevant and immediately useful for the description of the jumps that occur in first-order phase transitions. 

\subsection{Fluctuations}

In equiliubrium, the material has a behavior that, in a gross examination, looks time-independent. Hence many of the phenomena involved may be described by using time-averages of various quantities. This averaging is the basis of mean-field-theory techniques.   However, a more detailed look shows {\em fluctuations}, that is time dependence,  in everything.  These fluctuations will call for  additional calculational techniques beyond mean field theory, which will be realized with the renormalization group methods described in \ch{New}.  Here, I describe two important examples of fluctuations that arise near phase transitions.  
\subsubsection{Fluctuations I: Boiling}
In the process of ordering, typically a material will display large amounts of disorder.  For example, as the pressure is reduced at the liquid-gas coexistence line, a liquid turns into a vapor by an often-violent process of boiling.   The boiling produces bubbles of low-density vapor in the midst of the higher-density liquid. Thus the fluid, which is quite homogeneous away from its phase transition, shows a rapidly fluctuating density in its boiling region.  As every cook knows, one can reduce the violence of the fluctuations by making the boiling less rapid. Nonetheless,  it remains true that the fluid shows an instability in the direction of fluctuations in its region of boiling in the phase diagram.    

\subsubsection{Fluctuations II: Critical opalescence\la{opal}}  
A process, not entirely dissimilar to boiling, occurs in the equilibrium fluid near its critical point.   Observers have long noticed that, as we move close to the liquid-gas critical point, the fluid, hitherto clear and transparent, turns milky.  This phenomenon, called {\em critical opalescence}, was studied by   Marian Smoluchowski (1908)  and  Albert  Einstein (1910)\cite[p. 100]{Pais}.  Both recognized that critical opalescence was caused by the scattering of light from fluctuations in the fluid's density.  They pointed out that the total amount of light scattering was proportional to the compressibility, the derivative of the density with respect to pressure.  They also noted that the large amount of scattering near the critical point was indicative of anomalously large fluctuations in that region of parameters.    In this way, they provided a substantial explanation of critical opalescence\footnote{Einstein then used the explanation of this physical effect to provide one of his several suggested ways of measuring Avogadro's number, the number of molecules in a mole of material.}.  

\fig{susN} shows the divergence in the compressibility.
\subsection{Ornstein and Zernicke}

Leonard Ornstein and Frederik Zernike\cite{OZ}subsequently derived a  more detailed theory of critical opalescence. In modern terms,  one would say that the scattering is produced by small regions, droplets, of materials of the two different phases in the near-critical fluid.  The regions would become bigger as the critical point was approached, with the correlations extending over a spatial distance called the correlation or coherence length, $\xi$.    Ornstein and Zernike saw this length diverge on the line of coexistence between the two phases of liquid-gas phase transition as the critical point was approached in the form
\be
\xi = \text{constant} \times 1/|T-T_c|^\nu     \text{~~with~~}  \nu=1/2
\la{xi}
\ee
with $T-T_c$ being, once again,  the temperature deviation from criticality.    
Thus the correlation length does go to infinity as the critical point is approached.  This result is crucially important to the overall understanding of the critical point.   It is a realization of the extended-singularity-theorem idea that the phase transition, or more specifically criticality, reflects correlations over an infinite region of the system.

\subsection{Outline of paper} 
The next chapter defines the Ising model, a simple and basic model for phase transitions.   It then uses that model to describe the {\em extended singularity theorem}, which  describes the relationship between phase transitions, mathematical singularities in thermodynamic functions, and correlated fluctuations.  \ch{Mean} then defines mean field theories as one way of approaching the theory of phase transitions. The next chapter, \ch{1937}, describes the 1937 Landau theory as the pinnacle of mean field theory descriptions, but it is also the start of the replacement of that theory by fluctuation-dominated approaches.   In that same year, a conference in Amsterdam exhibited the confusion caused by the conflict between mean field theory and the extended singularity theorem.   The long series of studies that indicated a need for supplementing mean field theory is described in \ch{Beyond}.   Finally \ch{New} describes the development of a new phenomenology and  theory to understand fluctuations in phase transitions.

\section{The Ising model\la{IsingM}}
\subsection{Definition}
The Ising model is a conceptually simple representation of a system that can potentially show ferromagnetic behavior.  Its name comes from the physicist Ernst Ising\cite{Ising}, who studied it\footnote{Much of the historical material in this work is taken from the excellent book on critical phenomena  by Cyril Domb\cite{Domb}.} in conjunction with his adviser Wilhelm Lenz\cite{Lenz}.   Real ferromagnets involve atomic spins placed upon a lattice. The elucidation of their properties requires a difficult study via the band theory of solids. The Ising model is a shortcut that catches the main qualitative features of the ferromagnet.  It puts a spin variable  upon each site, labeled by $\bf{r}, $ of a simple lattice.  (See \fig{lattice}.)  Each spin variable, $\sigma_{\bf{r}}$, takes on values plus or minus one to represent the possible directions that might be taken by a particular component of a real spin upon a real atom.   

    The sum over configurations is a sum over all these possible values.   The Hamiltonian for the system is the simplest representation of the fact that neighboring spins interact with a dimensionless coupling strength, $K$, and a dimensionless coupling to an external magnetic field, $h$.  The Hamiltonian is given by
\be
-H/T=K  \sum_{nn} \sigma_r \sigma_s+ h \sum_r \sigma_r
\la{Ising}
\ee         
where the first sum is over all pairs of nearest neighboring sites, and the second is over all sites.  The actual coupling between neighboring spins, with dimensions of an energy, is often called $J$. Then $K=-J/ T.$  In turn, $h$ is proportional to the magnetic moment of the given spin times the applied magnetic field, all divided by the temperature.    Since lower values of the energy have a larger statistical likelyhood, the two term in \eq{Ising} reflect respectively a tendency of spins to line up with each other and also a tendency for them to line up with an external magnetic field. 
\begin{figure}
\begin{multicols}{2}
\includegraphics[height=6cm ]{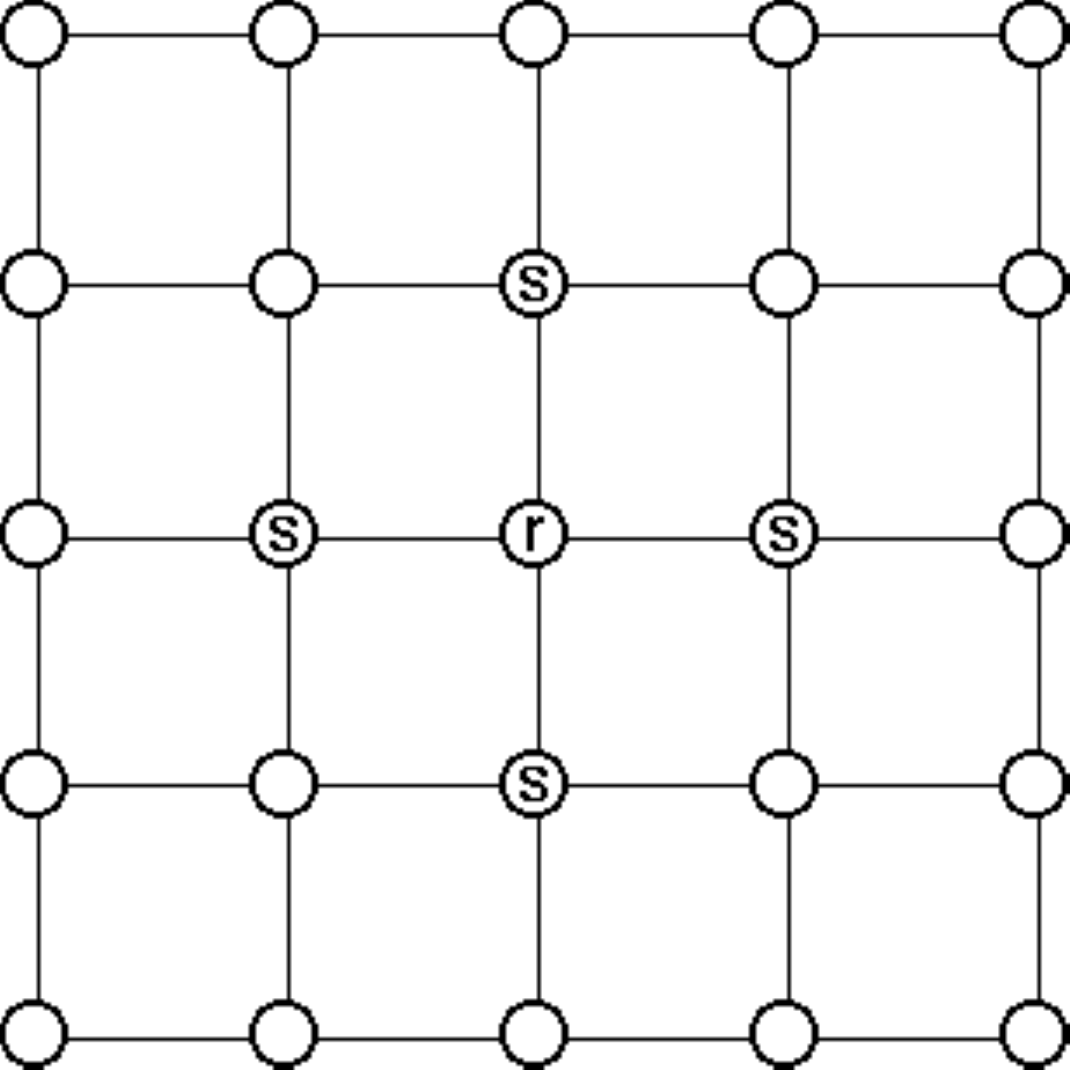}
\caption{
Lattice for two-dimensional Ising model.  The spins are in the circles.  The couplings, $K$, are the lines.  A particular site is labeled with an ``r''.  Its nearest neighbors are shown with an ``s''.    }
\la{lattice}
\end{multicols}
\end{figure}

\subsection{The extended singularity theorem} 
Particle spin is certainly a quantum mechanical concept.  There is no simple correspondence between this concept and anything in classical mechanics.  Nonetheless, the concept of spin fits smoothly and easily into the Boltzmann-Gibbs formulation of statistical mechanics.  When spins are present, the statistical sum in \eq{free} includes a quantum summation over each spin-direction.  In the Ising case, when the only variable is $\sigma$, standing for the $z$-component of the spin, that summation operation is simply a sum over the two possible values plus one and minus one of each spin at each lattice site.  

For the Ising model, the integral over configurations in \eq{free} is replaced by a sum over the possible spin values, thus making the result have particularly simple mathematical properties.    Take the logarithm of that equation and find:
\be
-F/T=\ln[\sum_c e^{-\mathcal{H}(c)/T}]
\la{freeln}
\ee 
On the right hand side of this equation one finds a simple sum of exponentials.  This is a sum of positive terms and it gives a result that is a smooth function of the parameters in each exponential, specifically the dimensionless spin-coupling, $K$, and the dimensionless magnetic field $h$.   A logarithm of a smooth function is itself a smooth function.    Therefore it follows directly that the free energy, $F$, is a smooth function of $h$ and $K$.

The reader will notice that this smoothness seems to contradict our definition of a phase transition, the statement that a phase transition is a singularity, i.e.,  failure of smoothness, in some thermodynamic quantity.  

This seeming contradiction is the key to understanding phase transitions.   No sum of a finite number of smooth terms can be singular.  However, for large systems, the number of terms in the sum grows quite rapidly with the size of the system.  When the system is infinite, the number of terms in infinite.  Then singularities can arise. Thus all singularities, and hence all phase transitions, are consequences of the influence of some kind of infinity.  Among the likely possibilities are infinite numbers of particles,  infinite volumes, or --more rarely--  infinitely strong interactions. Real condensed matter systems  often have large numbers of particles.  A cubic centimeter of air contains perhaps $10^{20}$ particles.   When the numbers are this large, the systems most often behave almost as if they had an infinity of particles.   

I am going to give a name to the idea that phase transitions only occur when the condensed matter system exhibits the effect of some singularity extended over the entire spatial extent of the system.  Usually the infinity arises because some effect is propagated over the entire condensed system, that is, over a potentially unbounded distance.  I am going to call this result the ``extended singularity theorem,'' despite the fact that the argument is rather too vague to be a real theorem.  It is instead a slightly imprecise mathematical property of real phase transitions.\footnote{Imprecision can often be used to  distinguish between the mathematician and the physicist. The former tries to be precise, the latter sometimes uses vague statements that can then be extended to cover more cases.  However, in precisely defined situations, for example the situation defined by the Ising model, the extended similarity ``theorem'' is actually a theorem\cite{Isakov}.}     

This theorem is only partially informative.   It tells us to look for a source of the singularity, but not exactly what we should seek.  In the important and usual case in which the phase transition is produced by the infinite size of the system, the theorem tells us that any theory of the phase transition should look to things that happen in the far reaches of the system.  What things?   How big are they? How should one look for them?  Will they dominate the behavior near the phase transition or be tiny?   The theorem is uninformative on all these points.     

Sometimes it is very hard to see the result of the theorem.   In an Ising or liquid-gas phase transition there is a singularity in the regions just touching the coexisting phases\cite{Fisher-L-Y,Andreev}. This singularity is very weak.  One must use indirect methods to observe or analyze it.   Conversely, near critical points,  singularities are very easy to observe and measure.   For example, in a ferromagnet,  the derivative of the magnetization with respect to applied magnetic field, is infinite at the critical point.  (See \fig{susN}.)   

By looking at simulations of finite-sized Ising systems one can see how the infinite size of the system enters the susceptibility.    \fig{susN} is a set of plots of susceptibility  versus temperature in an Ising system with a vanishingly small positive magnetic field.  The different plots show what happens as the number of particles increases toward infinity.   As you can see, the finite $N$ curves are smooth, but the infinite-$N$ curve goes to infinity.  This infinity is the singularity.    It does not exist for any finite value of $N$. However, as $N$ gets larger,  the finite-$N$ result approaches the infinite-$N$ curve.   When we look at a natural system, we tend to see phase transitions that look very sharp indeed, but are actually slightly rounded.   However, a conceptual understanding of phase transitions requires that we consider the limiting, infinite-$N$, case.

\begin{figure}
\begin{multicols}{2}
\includegraphics[height=7cm ]{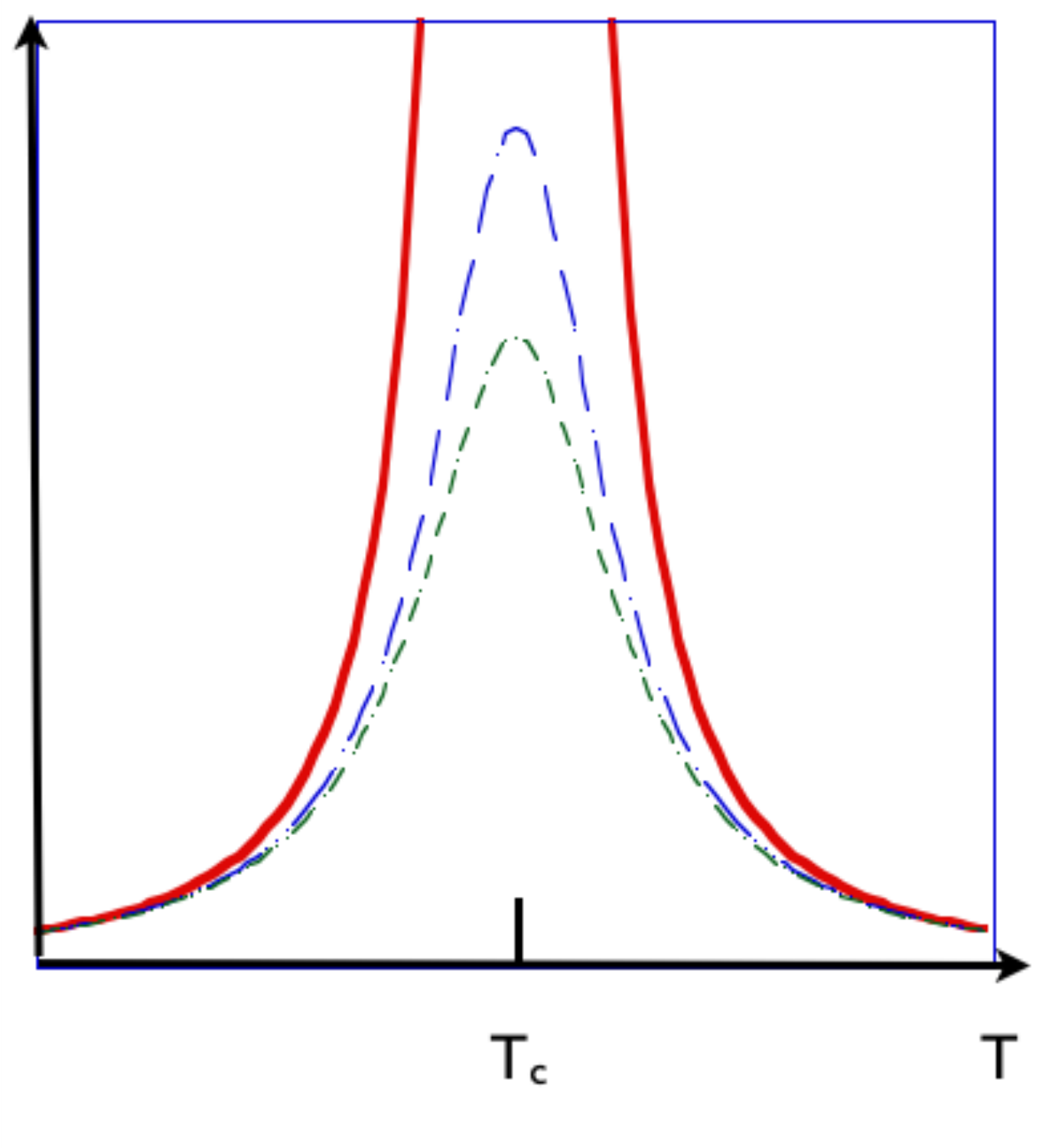}
\la{susc}
\caption{Cartoon view of a singularity in a phase transition.  The {\em magnetic susceptibility}, the derivative of the magnetization with respect to magnetic field,  is plotted against temperature for different values of $N$. The thick solid curve is shows  the susceptibility in an infinite system.  The dashed curves apply to systems with finite numbers of particles, with the higher line being the larger number of particles.    The compressibility of the liquid-gas phase transition also shows this behavior.     }
\la{susN}
\end{multicols}
\end{figure}

\fig{ferro} is the phase diagram of the Ising model.  The $x$ axis is the magnetic field; the y axis is the temperature.   This phase diagram  applies when the lattice is infinite in two or more dimensions. There is no phase transition for lower dimensionality.  

\begin{figure}
\begin{multicols}{2}
\includegraphics[height=6cm ]{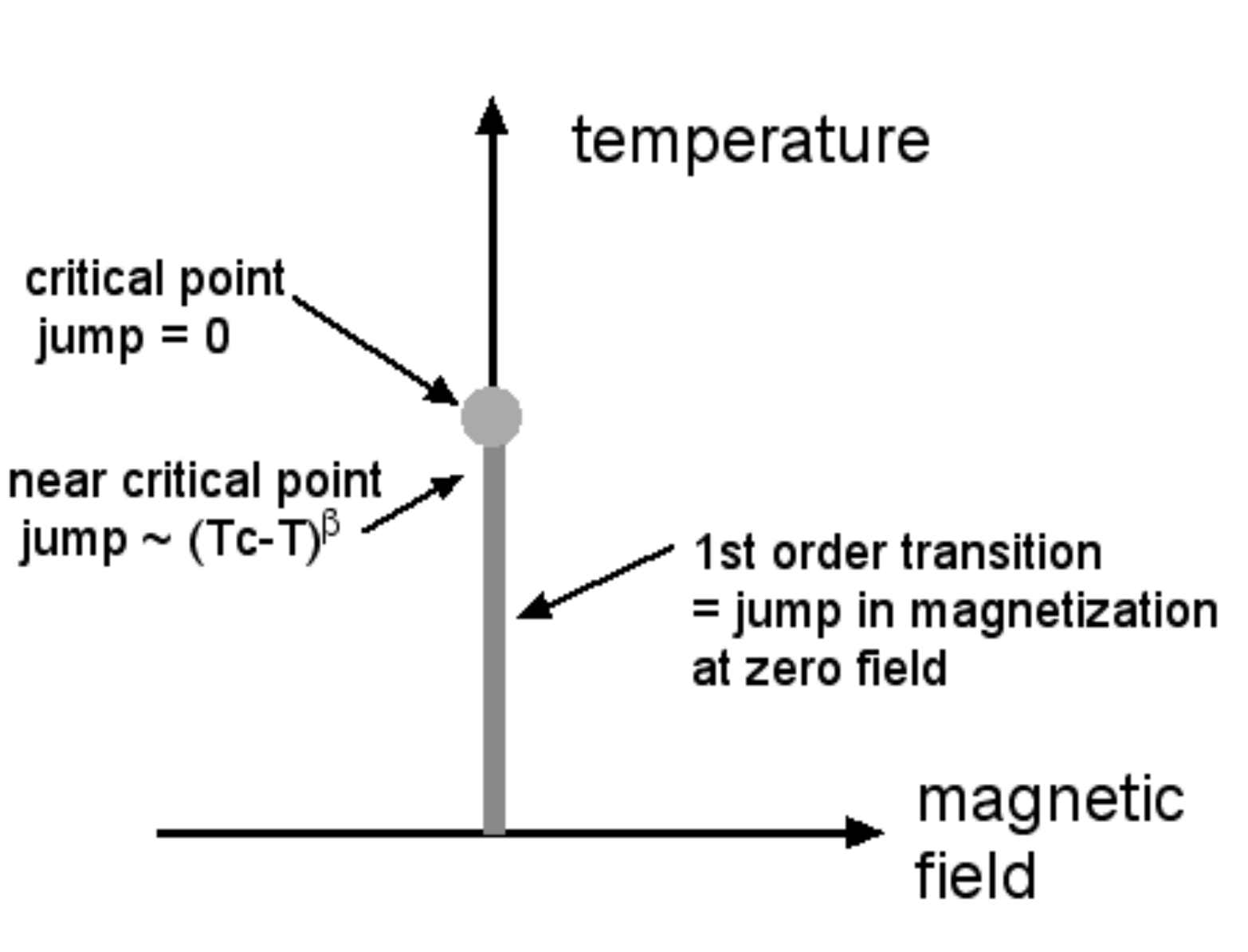}
\caption{Phase diagram for ferromagnet and Ising model.   The jump in magnetization occurs at zero magnetic field.  In this representation, the jump region has been reduced to a line running from zero temperature up to the critical point.  }
\la{ferro}
\end{multicols}
\end{figure}

\section{More is the same\la{Mean}}
This chapter describes mean field theory, which forms the basis of much of modern many-particle physics and field theory.  So far, we said that an infinite statistical system sometimes has a phase transition, and that transition involves a discontinuous jump in a quantity called the order parameter.   But we have given no indication of how big the jump might be, nor of how the system might produce it.   Mean field theory provides a partial and approximate answer to that question.

We begin with the statistical mechanics of one spin in a magnetic field. Then, we extend this one-spin discussion to describe how many spins work together to produce ferromagnetism.

\subsection{One spin}
A single spin in a magnetic field can be described by a simplification  of the Ising Hamiltonian of \eq{Ising},
$
- H/T = h  \sigma.
$ 
As before, $\sigma$  is a component of the spin in the direction of the magnetic field. This quantum variable takes on two values $\pm 1$, so that probability distribution  of \eq{MB} gives the average value of the spin as
\be
<\sigma>   = \sum_{ \sigma=\pm 1} ~\sigma~ e^{h\sigma-F}=\tanh h
\la{av}
\ee
(In general, we write the statistical average of any quantity, $q$,  as $<q>$. )

\subsection{Many spins; mean fields\la{Many}}
The very simple result, \eq{av}, appears again when one follows Pierre Curie\cite{Curie} and Pierre Weiss\cite{Weiss} in their development of a mean field theory of ferromagnetism.  Translated to the Ising case, their theory  would ask us to concentrate our attention upon one Ising variable, say the one at $\bf{r}$.   We would then notice that this one spin would  see a field with the value 
\be
h({\bf r}) +K \sum_{{\bf s}~nn~to~ {\bf r}}  \sigma_{\bf s}
\la{environ}
\ee
where $h({\bf r})$ is the dimensionless magnetic field at ${\bf r}$ and the sum covers all the spins with positions, ${\bf s}$, sitting at nearest neighbor sites to ${\bf r}$.   To get the mean field result, replace the actual values of all the other spins, but not the one at  $\bf{r}$, by their average values and find, by the same calculation that gave \eq{av}  a result in which the average is once more
\bsubs  \la{mft} 
\be
<\sigma_{\bf r}> = \tanh h^{\text{eff}}({\bf r}) 
\la{avmft}
\ee
 but now the actual field is replaced  by an effective field
\be
h^{\text{eff}}({\bf r})=h({\bf r}) +K \sum_{{\bf s}~nn~to~ {\bf r}} < \sigma_{\bf s} >
\la{eff}
\ee
\esubs

\subsection{Mean field results}
Given this calculation of basic equations for the local magnetization, $<\sigma_{\bf r}>$, we can go on to find many different aspects of the behavior of this mean field magnet.  We notice that when $h$ is independent of position, the equation for  $<\sigma>$ has a critical point, i.e., an ambiguity in its solution,  at zero magnetic field and $Kz=1$.  When we expand the equations around that critical point, we can find that the magnetization obeys  a cubic equation like that of the van der Waals theory (or equally  the Landau theory as described below in \se{Landau}).  Thus, there is a full and complete correspondence between the van der Waals theory of the liquid-gas transition and the ferromagnetic  mean field theory near its critical point, as one can see by comparing   \fig{PVT} with \fig{ferro}.

A brief calculation using the mean field theory gives the magnetic susceptibility, the derivative of the magnetization with respect to magnetic field,  to  diverges as $1/|T-T_c|$ near the critical point.  The analogy just mentioned between the liquid-gas system and the magnetic phase transition makes this magnetic susceptibility the direct analog of the compressibility, that had an infinity which was used by Einstein to explain critical opalescence. (See \se{opal})  

On the other hand, Ornstein and Zernike calculated the fluid analog of the more disaggregated quantity
\be
g(\mathbf{r,s}) =  \frac{\partial <\sigma_\mathbf{r}>}{ \partial   h(\mathbf{s})}
\ee called the spin correlation function. 
A sum over all lattice sites, $\bf{s,}$ of this correlation function will give the susceptibility. Then $g$ can be evaluated from the equations of mean field theory\cite[page 232]{LPK00} as
\be
g(\mathbf{r,s})=z ~a~\frac{e^{-|\mathbf{r-s}|/\xi}}{4\pi|\mathbf{r-s}|}
\la{gsoln}
\ee
in the simplest case: three dimensions, $h=0$, $t$ small but greater than zero, and separation distance large compared to the lattice spacing, $a$.  In \eq{gsoln}, $\xi$ is the correlation length  that describes the range of influence of a change in magnetic field. Its value, given by \eq{xi}, shows that the correlation length diverges as critical is approached.  This behavior is an expected consequence of the extended singularity theorem, which asks for infinite ranges of influence at phase transitions.  

We previously argued that the extended singularity theorem called for fluctuations extending over large distances.  Indeed that call is precisely answered by \eq{gsoln} and \eq{xi}.  A theorem of statistical mechanics relates the correlation function to spin fluctuations via
\be
 \la{fluctu}
g(\mathbf{r,s}) =  < \sigma_\mathbf{r}\sigma_\mathbf{s}>-  < \sigma_\mathbf{r}>< \sigma_\mathbf{s}> 
=< (\sigma_\mathbf{r}-<\sigma_\mathbf{r}>)(\sigma_\mathbf{s}-<\sigma_\mathbf{s}>)>
\ee

Note the scaling of the spin correlation function.  For relatively small values of the distance, the correlation function in \eq{gsoln} has a form in which $g$ varies as one divided by the distance.  It is conventional to describe this correlation function, varying as $|\mathbf{r-s}|^{-2x}$, by saying that there are two local quantities contributing to the correlation  and then saying that each scales as distance to the power $x$.   Therefore in this case, the index going with the order parameter is $x=1/2$.

\subsection{Representing critical behavior by power laws }
The reader will, no doubt, have noticed the appearance of ``power laws'' in the description of behavior near critical points. In these laws, some critical property is written as a power of a quantity that might become very large or very small, as for example, magnetization =  constant  $\times~  t^\b$.    So far, we have seen  laws like this in the behavior of the order parameter(\eq{beta}), the correlation length(\eq{xi}), the magnetic susceptibility(\fig{susc}), and the correlation function(\eq{gsoln}).  Why does this power function appear repeatedly? 

All of this singular behavior is rooted in the fact that phase transitions produce a variation over a tremendous range of length scales.  For example, the basic interactions driving most phase transitions occur on a length scale described by the distance between atoms or molecules, that is some fraction of a nanometer ($10^{-9}$ meters). On the other hand, we observe and work with materials on a characteristic length scale of centimeters ($10^{-2}$ meters).   The crucial issue in phase transitions is how does the material interpolate phenomena over this tremendous length scale.   The answer is roughly speaking that all the physical quantities mentioned follow the changes in the length scale.  

As we shall see in \se{Wilson}, in renormalization calculations, the changes of the length scale in turn follows from multiplicative laws. To get to a tremendous change in length scale, $\ell$, one puts many small steps $\ell_1, \ell_2, \ell_3, \cdots$ into the renormalization calculation   and the big change is produced by the multiplication of these factors, 
$$\ell=\ell_1 \times \ell_2 \times\ell_3 \times \cdots. $$   This kind of behavior is explicitly built into renormalization calculations\footnote{Calculations of the effects of scale changes are much more implicit in mean field theories than in renormalization theories. In both cases we are treating variation over a huge range of  scales, and power laws are a likely way of describing this huge range of variation.   However, because the mean field theories deal less directly with scale transformations they do not get the relation between  the renormalization scalings of fluctuations and free energy quite right.    }.  

Scale transformation is a symmetry operation.  It describes an underlying symmetry of nature in which every scale-- kilometer, centimeter, nanometer-- is equally good for describing nature's basic laws. Whenever a physical phenomenon reflects a symmetry operation, observed physical quantities must be mathematical representations of that symmetry operation.   That is why we use scalars, vectors and tensors to describe quantities that obey, say, the usual rotational symmetry. The same thing works for scale transformations.  Here, power laws reflect the symmetries built into multiplication operations.   The physical quantities behave as powers, $\ell^x$, where $x$ can be rational or irrational, positive or negative, or indeed zero.  In the last case, the limiting behavior is that of a logarithm instead of a power, as is actually obtained in the heat capacity of the Onsager solution\cite{Onsager} of the two-dimensional Ising model.    

The wide range of length scales also applies in particle physics where the basic scales for  interactions may be vastly different from the scale at which observations are performed.  Thus, in particle physics it is also true that renormalization and scaling have to interpolate behaviors over very large length scales.   

Whenever one has a power law, say $<\sigma> =(-t)^\b$, one has an power, here $\b$.   This power is called a ``critical exponent'' or a ``critical index''.   During the many years in which critical behavior has been a subject of scientific study, many human-years of scientific effort has been devoted to the accurate determination of these indices.  Sometimes scientists complained that this effort was misplaced.  After all, there is little insight to be obtained from the statement that $\b$ (the index that describes the jump in the liquid-gas phase transition)  has the value 0.31 versus 0.35 or 0.125 or 0.5.   But these various values can be obtained from theories that give a direct calculation of critical quantities or related them one to another.  The calculations or relations come from ideas with considerable intellectual content. Finding the index-values then gave an opportunity to check the theory and see whether the underlying ideas were sound.  Thus the small industry of evaluating critical indices supports the basic effort devoted to understanding critical phenomena.

\section{The Year 1937: A Revolution Begins\la{1937}}
\subsection{Landau's generalization\la{Landau}}
Lev Landau followed van der Waals, Pierre Curie, and Ehrenfest  in noticing a deep connection among different phase transition problems\cite{Daugherty}.   Landau translated this observation into a mathematical theory in a novel and interesting way.    Starting from the recognition that, in the neighborhood of a critical point,  each phase  transition was a manifestation of a broken symmetry,  he used the order parameter to describe the nature and  the extent of  symmetry breaking\cite{Landau}.

Landau generalized the work of others by writing the free energy as an integral over all space of an appropriate function of the order parameter. The dependence upon ${\bf r}$ indicates that the order parameter is considered to be a function of position within the system.  In the simplest case, described above, the phase transition is one in which the order parameter, say the magnetization, changes sign.\footnote{The symmetry of the phase transition is reflected in the nature of the order parameter, whether it be a simple number (the case discussed here), a complex number (superconductivity and superfluidity), a vector (magnetism), or something else.}  In that case, the appropriate free energy takes the form
\be
-F/T= \int ~ d{\bf r} \big[  A h({\bf r}) \Psi({\bf r})  +B~ \Psi({\bf r})^2 + C ~\Psi({\bf r})^4 +  D~[\nabla \Psi({\bf r}) ]^2 +\cdots       \big]
\la{F}
\ee
where $A,B,C, \ldots $ are parameters that describe the particular material and $\Psi({\bf r})$ is the order parameter at spatial position ${\bf r}$.     In recognition of the delicacy of the critical point, each term goes to zero more rapidly than $\Psi({\bf r})^2$ as criticality is approached.

The next step is to use the well-known rule of thermodynamics that the free energy is minimized  by the  achieved value of every possible macroscopic thermodynamic variable within the system.   
Landau took the magnetization density at each point to be a thermodynamic variable that could be used to minimize the free energy.   
Using the calculus of variations one then gets an equation for the order parameter: 
\be
0=  h({\bf r}) +2(B/A)~ \Psi({\bf r})  +4  (C/A) ~\Psi({\bf r})^3- 2  (D/A)~\nabla^2 \Psi({\bf r}) 
\la{deltaF1}
\ee
One would get a result of precisely this form by applying the mean field theory magnetization equation near the critical point.     The $B$-term is identified by this comparison as being proportional to the temperature deviation from criticality, $B= - A t/2$.

In some sense, of course, Landau's critical point theory is nothing new.  All his results are contained within the earlier theories of the individual phase transitions.   However, in another sense his work was a very big step forward.   By using a single formulation that could encompass all critical phenomena with a given symmetry type, he pointed out the close similarity among different phase transition problems.   And indeed in the modern classification of phase transition problems\cite{67Review} the two main elements of the classification scheme are the symmetry of the order parameter and the dimension of the space.  Landau got the first one right but not, at least in this variational formulation, the second classifying feature.       On the other hand, Landau's inclusion of the space gradients that then  brought together the  theory's  space dependence and its thermodynamic behavior also seems, from a present-day perspective, to be right on.

\subsection{Summary of mean field theories}
As already mentioned, Landau's 1937 result provides a kind of mean field theory that agrees in all essential ways with the results of the main previous workers.  The only difference is that Landau produced a specialized theory  intended to apply mostly to the region near the critical point.  From the point of view of the discussion that will follow the main points of his theory and the earlier ones are: \begin{itemize}
\item Universality.  The Landau theory gives an equation for the order parameter as a function of the environmental parameters that is universal:  It only depends upon the kind of symmetry reflected in the ordering. 
\item Symmetry.    A first-order phase transition is often, but not always, a reflection of a change in the basic symmetry of the condensed system.
\item  Interactions.  This symmetry change is usually caused by local interactions among the basic constituents of the system.   
\item Scaling.   The results depend upon simple ratios of the environmental parameters raised to  powers.  For example in the ferromagnetic transition all physical quantities depend upon the ratio $t^{3}/h^{2}$.  In subsequent theories, the restriction to simple powers will disappear.        
\item Order parameter jump.   There is a discontinuous jump in the order parameter at the first-order phase transition as in \eq{beta}.  The jump goes to zero as the critical point is approached, with critical index $\b=1/2$.
\item Correlation length.   The correlation length goes to infinity at criticality as in \eq{xi} with an index $\nu=1/2$. 
\end{itemize}
As we shall see, for many purposes, the mean field theories have been replaced by a {\em renormalization group} theory of phase transitions.   The  qualitative properties of mean field theory, like universality and scaling, have been retained.  On the other hand,  all the quantitative properties of the theory, for example, the values of the critical indices,  have been replaced.

\subsection{Away from corresponding states--- toward universality}
Landau's calculation represented the high-water mark of the class of theories described as ``mean field theories''.   He showed that all of them could be covered by the same basic calculational method.   They differed in the symmetries of the order parameter, and different symmetries could give different outcomes.  However, within one kind of symmetry the result was always the same.   This uniform outcome was very pleasing for many students of the subject, particularly so for the physicists involved.    We physicists especially like mathematically based generalizations and Landau had developed an elegant generalization, which simplified a complex subject. 

However, Landau's uniformity was different from the theoretical idea of uniformity that had come before him.  Earlier work had been based upon the idea that different fluids have almost identical relations between their pressure temperature and density.  This idea is called the ``principle of corresponding states.''  

This principle of corresponding states had broad support among the scientists working on phase transitions. Starting with van der Waals,  continuing with the work of Einstein\cite[p. 57]{Pais}, George Uhlenbeck, and  E. A. Guggenheim\cite{Guggenheim}, work on phase transitions was inspired by the aim and hope that the phase diagrams of all fluids would be essentially alike.   However, Landau's work marked a new beginning.  His method would apply only near a critical point and his version of corresponding states could be expected to apply only in this region.  So Landau  deepened the theory  but implicitly also narrowed its domain of application to a relatively small region of the phase diagram.   This was the start of a new point of view, which we shall see develop in the rest of this paper.    The new point of view would come with a new vocabulary so that instead of corresponding states people would begin to use the word ``universality\cite{me}.''

\subsection{Statistical confusion: A meeting in the Netherlands}
The extended singularity theorem (See \ch{IsingM}.) presents both an opportunity and a challenge for understanding phase transitions.  The theorem is self-evident in the case of the Ising model with its simple sums and exponentials.  It is less obviously true for the statistical mechanics of the liquid-gas phase transition since, in this case, the calculation of the free energy includes integrals and also unbounded potentials.  However, the theorem remains true for that transition.  Thus the theorem would then demand an infinite system for a sharp liquid-gas phase transition.  On the other hand, the van der Waals mean field argument would, for example, give a sharp phase transition in small systems.   This contradiction might serve as a confusing element in the development of a theory of phase transitions.     

The contradiction was as old as the first definitions of statistical mechanics and phase transitions, but was apparently not discussed for many years.  It might well, however, have come up at a  1937 meeting held in  Amsterdam to celebrate the centenary of van der Waals' birth.   Hendrik Kramers, George Uhlenbeck, and Peter Debye were all present at that occasion. According to  Uhlenbeck\cite{Uhlenbeck},  at that meeting Kramers pointed out that the sharp singularity of a phase transition could only occur in a system with  some infinity built in and, for that, that an infinite system is required.  Then the van der Waals theory's prediction of a phase transition in a finite system could be viewed as a grave failure of mean field theory and maybe even of statistical mechanics.   E. G. D.  Cohen's described material by Uhlenbeck\cite{Cohen}:  ``Apparently the audience at this van der Waals memorial  meeting in 1937, could not agree on the above question, whether the partition function could or could not explain a sharp phase transition. So the chairman of the session, Kramers, put it to a vote."   According to the story I heard, again based upon the recollection of Uhlenbeck, after some discussion the meeting voted on the proposition ``Can statistical mechanics describe the liquid state.''     The meeting is said to have split 50-50, with Debye (!) voting no! 
  
Clearly half the people at that meeting were wrong.  Seventy plus  years later one can see the right answer, in close analogy to our understanding of irreversibility.   Infinite size is required for a sharp phase transition, but a large system can very well approximate the behavior of the infinite system.   Finite size  slightly rounds off and modifies the sharp corners shown in the plot of \fig{PVT}.  Conventional statistical mechanics, following the path begun by Andrews, van der Waals, and Maxwell, can describe quite well what happens in the liquid region, especially if one stays away from the critical region and from boiling.  In fact, there are theories, including one by John Weeks, David Chandler, and Hans Anderson\cite{WCA}, that do a good job of describing the liquid region of the fluid. 

However, the extended singularity theorem does have its effects.  There are indeed  singularities near the first-order transition\cite{Fisher-L-Y,Andreev}. These singularities are very weak in the Ising and liquid-gas transition, but will be stronger when an unbroken symmetry remains after the symmetry breaking of the first-order transition. (We see such a residual symmetry in the Heisenberg model of ferromagnetism. ) Also, there are quite strong singularities in the neighborhood of the critical point, not correctly described by mean field theory.All these singularities are consequences of fluctuations, which are not included in the mean field approach.

However, the Amsterdam meeting was quite right, in my view, to be disquieted by the applicability of statistical mechanics.  But they focused upon the wrong part of the phase diagram. The liquid region is described correctly by statistical mechanics. But this theory does not work well  in the two-phase, ``boiling'', region of \fig{PVT}.    Here the fluctuations entirely dominate and the system sloshes between the two phases.  The behavior of the interface that separates the phases is determined by delicate effects of dynamics and previous history, and by hydrodynamic effects including gravity, surface tension, and the behavior of droplets.  Hence the direct application of statistical mechanics is fraught with difficulty precisely in the midst of the phase transition.  Thus, the extended singularity theorem suggests that a new theory is required to treat all the fluctuations appearing near singularities.

\section{Beyond (or Beside) Mean Field Theory\la{Beyond}}
Weaknesses of mean field theory began to become apparent to the scientific community immediately after Landau's statement of the theory in its generalized  form.  This chapter will describe the process of displacement of mean field theory, at least for behavior near the critical point,   which we might say began in 1937 and culminated in Kenneth Wilson's enunciation of a replacement theory in 1971\cite{Wilson}.  

Landau's theory provided a standard and a model for theories of general phenomena in condensed matter physics.  Looking at Landau's result one might conclude that a theory  should be as general and elegant as the phenomena it explained.   The mean field theories that arose before (and after) Landau's work were partial and incomplete, in that each referred to a particular type of system.   That was certainly necessary in that the details of the phase diagrams were different for different kinds of systems, but somewhat similar for different materials of the same general kind.  Landau's magisterial work swept all these difficulties under the rug and for that reason could not apply to the whole phase diagram of any given substance.   Thus, if Landau were to be correct he would most likely be so in the region near criticality. Certainly his theory is based upon an order parameter expansion that only is plausible in the critical region.   However, precisely in this region, as we shall outline below, both theoretical and experimental facts contradicted his theory.

\subsection{Experimental facts}

The ghosts of Andrews and van der Waals might  have whispered to Landau  that a theory that predicts $\b=1/2$ near criticality cannot be correct.   However  a much larger body of early work on fluids had pointed to this conclusion.  These early data, developed and published by J.E. Verschaffelt\cite{Verschaffelt} and  summarized by Levelt Sengers\cite{Levelt-Sengers}, touched almost every aspect of the critical behavior of fluids.  Verschaffelt particularly stresses the incompatibility of the data with mean field theory.    

These same experimental facts appear once more in the 1945 work of E.A. Guggenheim\cite{Guggenheim}, who compared data  for a wide variety of fluids.    He says ``The principle of corresponding states may safely be regarded as the most useful by-product of van der Waals' equation of state.    While this [van der Waals] equation of state is recognized to be of little or no value, the principle of corresponding states as correctly applied is extremely useful and remarkably accurate.''   He examined data  for seven fluids on the line of the liquid-gas phase transition, 
and fit the data to a power law with $\b=1/3$, rather than the mean field value $\b=1/2$. The latter value clearly does not work; the former fits reasonably well. Thus ``corresponding states'' receives support in this region, but not mean field theory {\em per se}.   

But neither Guggenheim nor Heike Kamerlingh Onnes\cite{Levelt-Sengers} before him was ready to receive information suggesting that behavior in the critical region was special, so that the former rejected mean field theory while the latter accepted it with reservations as to its quantitative accuracy.

Later,  near-critical data on heat capacity, the derivative of average energy with respect to temperature,   became available. Mean field theory predicts a discontinuity in the constant volume heat capacity as in \fig{cv}.    L. F. Kellers\cite{Kellers} looked at the normal fluid to superfluid transition in helium-4.  (See \fig{cvs}.) The data on this phase transition seemed to support the view that the heat capacity diverges weakly, perhaps as a logarithm of $|T-T_c|$, as criticality is approached.    Similar heat capacity curves were observed by Alexander Voronel'\cite{Voronel,Voronel1} in the liquid-gas transition of classical gases, and in further work in helium\cite{ML}.

\begin{figure}
\begin{multicols}{2}
\includegraphics[height=4cm ]{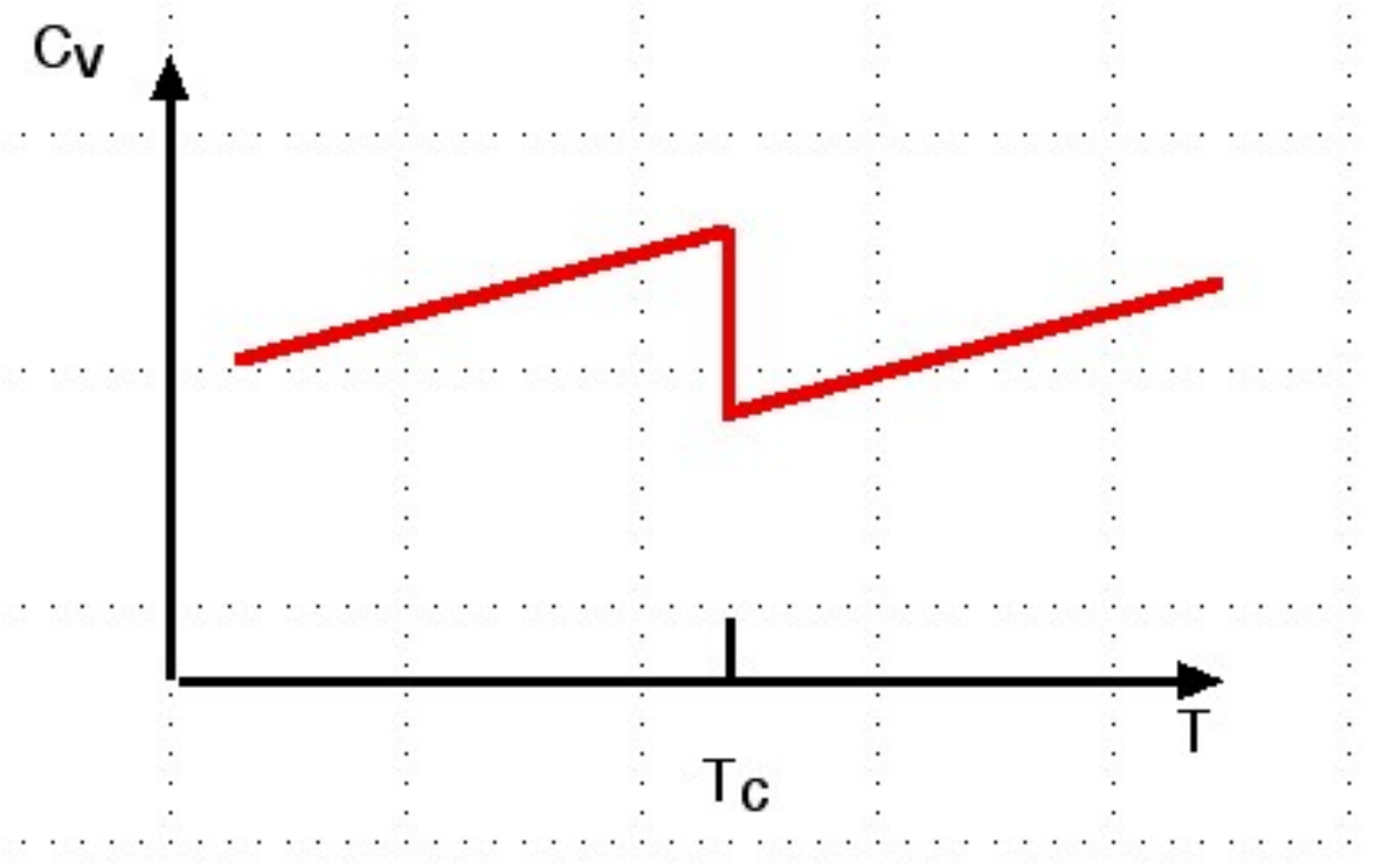}
\caption{Cartoon sketch of heat capacity in neighborhood of critical temperature as predicted by mean field theory.   The heat capacity is higher below $T_{c}$  because there is an additional temperature dependence in the free energy in this region produced by a term proportional to the square of the order parameter.   }
\la{cv}
\end{multicols}
\end{figure}   

\begin{figure}
\includegraphics[height=8cm ]{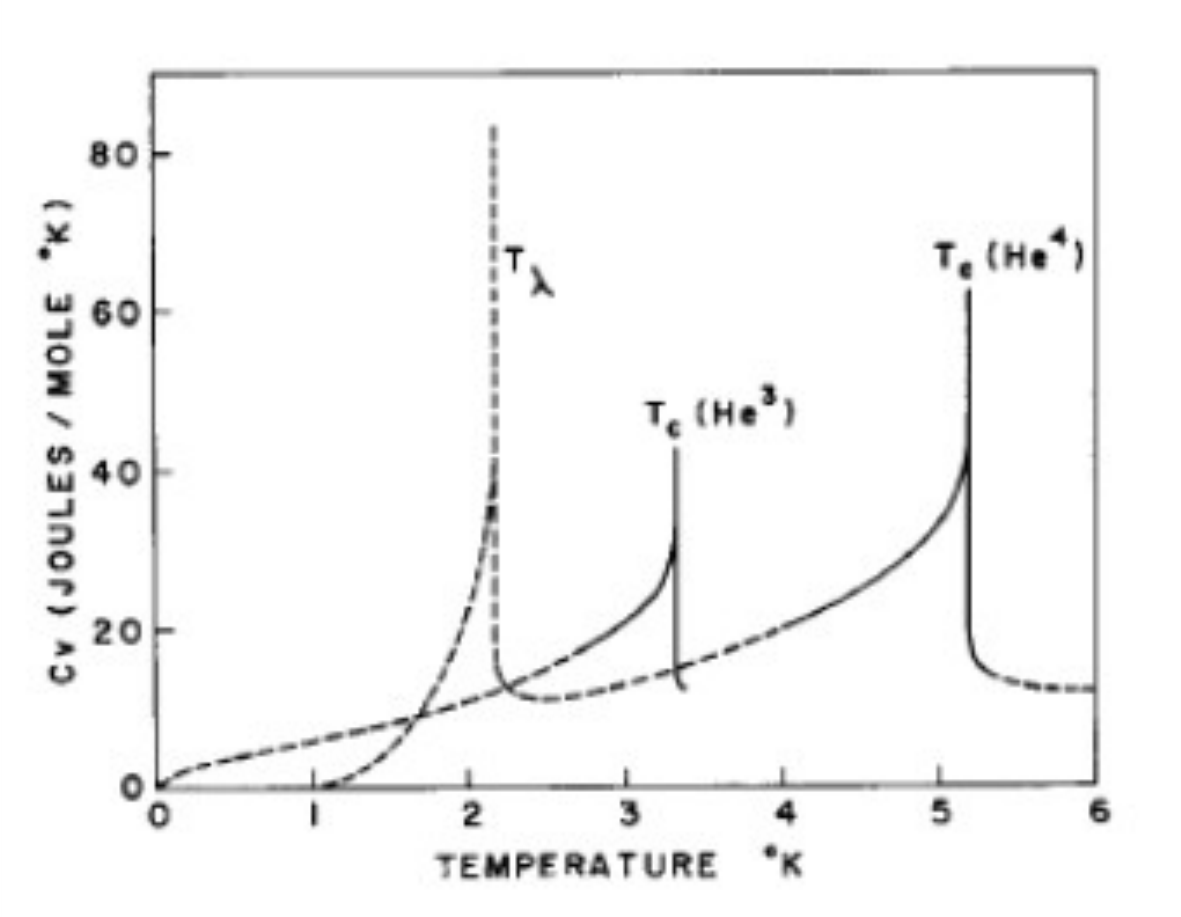}
\caption{Heat capacity as measured.  This picture,  the work of Moldover and Little\cite{ML}, shows measured heat capacities for the normal-superfuid transition of helium-4, labeled as $T_{\lambda}$, and the liquid-vapor transition of helium-3 and helium-4.   Note that all three heat capacities seem to go to infinity, in contrast to the prediction of mean field theory. These data certainly draw one's attention to the critical region. }
\la{cvs}
\end{figure}  

\subsection{Theoretical facts}
As we have seen, experimental evidence suggested that mean field theory was incorrect in the critical region.    A further  strong argument in this direction came from Lars Onsager's  exact solution\cite{Onsager} of the two-dimensional Ising model, followed by C.N. Yang's calculation\cite{Yang} of the zero-field magnetization for that model.   Onsager's  result for the heat capacity diverged as the logarithm of $T-T_c$ as did the experimental observations, as shown for example in \fig{cvs},  but did not resemble the discontinuity of mean field theory.\footnote{Onsager's results looked different from those depicted in \fig{cvs} in that they showed much more symmetry between the high temperature region and the low temperature region. This difference reflects the fact that two-dimensional critical phenomena are markedly different in detail from three-dimensional critical phenomena. Further, subsequent work has indicated that none of the heat capacity singularities shown in \fig{cvs} are actually logarithmic in character.  They are all power law singularities.  }

Yang's results, for which $\b=1/8$, also disagreed with mean field theory, which has $\b=1/2$.  
The Onsager solution implies a correlation length  with $\nu=1$,
which is not the mean field value $\nu=1/2$; see \eq{xi}. 

The most systematic theoretical discrediting of mean field theory came from the series expansion work of the King's College (London) school, under the leadership of Cyril Domb, Martin Sykes, and --after a time-- Michael Fisher\cite{Domb,Niss}.  Recall that the Ising model is a simplified model that can be used to describe magnetic transitions. It is described by a strength of the coupling between neighboring spins proportional to a coupling constant $J$.  The statistical mechanics of the model is defined by the ratio of coupling to temperature, specifically $K=-J/T$.   One can get considerable information about the behavior of these models by doing expansions of quantities like the magnetization and the heat capacity in power series in $K$, for high temperatures, and $e^{-K}$ for low temperatures.   The group at King's developed and used methods for doing such expansions and then analyzing them to obtain approximate values of critical indices like $\b$ and $\nu$.   The resulting index-values in two dimensions  agreed very well with values derived from the Onsager solution.  In three dimensions, models on different lattices gave  index values roughly agreeing with experiment on liquids and magnetic materials, but differing substantially from predictions of mean field theory.   This work provided a powerful argument indicating that mean field theory was wrong, at least near the critical point.   It also played a very important role in focusing attention upon that region. 

Another reason for doubting mean field theory, ironically enough, came from Landau himself.   In 1941, Andrei Kolmogorov\cite{K41} developed a theory of turbulence based upon concepts similar to the ones used in mean field theory, in particular the idea of a typical velocity scale for velocity differences over a distance $r$.  These differences would, in his theory, have a characteristic size that would be a power of $r$.  Landau criticized Kolmogorov's theory saying that it did not take into account fluctuations\cite{Frisch}, whereupon Kolmogorov modified the theory to make it substantially less similar to mean field theory\cite{K62}.  

\subsection{Spatial structures}
The spatial structure of mean field theory does not agree with the theorem that phase transitions can only occur in infinite systems.  Mean field theory is based on the alignment of order parameter  values at neighboring sites, so that particles will order if neighboring particles are ordered also.   Any collection of coupled spins can have a mean field theory phase transition.  Thus, two spins and a bond are quite sufficient to produce a phase transition in a mean field argument like that in \se{Many}.  On the other hand, the extended singularity theorem insists that the occurance of a phase transition  requires some sort of infinity, most often the existence of an infinite number of interacting parts within the system.  

As we shall see, what is wrong with mean field theory is that in the critical region the effect of the average  behavior of the order parameter can be completely swamped by fluctuations in this quantity.  In 1959 and 1960, A. P. Levanyuk\cite{Levanyuk} and Vitaly Ginzburg\cite{Ginzburg} described a criterion that one could use to determine whether the behavior near a phase transition was dominated by average values or by fluctuations.  For example, when applied to critical behavior of the type seen in the simplest version of the Ising model, this criterion indicates that fluctuations dominate in the critical region whenever the dimension  is less than or equal to four.   Hence mean field theory is wrong\cite{67Review} for all the usual  critical phenomena in systems with dimension smaller than or equal to four\footnote{There are exceptions.  Mean field theory works quite well whenever the forces are sufficiently long-ranged so that many different particles will interact directly with any given particle.  By this criterion mean field theory works well  for the usual superconducting materials studied up through the 1980s\cite{BCS,BCS1}, except extremely close to the critical point.   However, mean field theory does not work for the newer ``high-temperature superconductors'', a class discovered in 1986 by Georg Bednorz and Alexander M\"{u}ller\cite{MB}.}.    Conversely,  this criterion suggests that mean field theory gives the leading behavior above four dimensions.

\section{New Foci; New Ideas\la{New}}
\subsection{Bureau of Standards conference}
So far, the field of phase transitions had lived up perfectly to Thomas Kuhn's\cite{Kuhn} view of the conservatism of science. Before World War II, the only theory of phase transitions was mean field theory. No theory or model yeilded  \eq{beta} with any value of $\b$ different from one half.  There was no focus for anyone's discontent.  For this reason, the  mean-field-theory point of view continued  on, despite evidence to the contrary, until a set of events occurred that would move the field in a new direction.   One crucial event was the  conference on critical phenomena held at the U.S. National Bureau of Standards in 1965\cite{NBS}. The late Melville Green was the moving spirit behind this meeting.  The point of this conference was that behavior near the critical point formed a separate body of science that might be studied on its own merits, independent of the rest of the phase diagram.   In the years just before the conference, enough work\cite{DM,FisherReview,HellerReview} had been done so that the conference could serve as an inauguration of a new field.    We have mentioned the experimental studies of Kellers and of Voronel's group.  At roughly the same time important theoretical work was done by Alexander Patashinskii and Valery Pokrovsky\cite{PP}, Benjamin Widom\cite{Widom,WidomII} and myself\cite{LPK1966}, which would form a basis for a new synthesis.  The experimental and theoretical situation just after the meeting was summarized in reviews\cite{67Review,HellerReview,FisherReview}. This section begins by reporting on those new ideas and then describes their culmination in the work of Kenneth G. Wilson\cite{Wilson}.

\subsection{Correlation function calculations}
For many years the Landau group had been using field theory to describe the critical point.   Two young theoreticians, Patashinskii and Pokrovsky,  focused their attention upon  the correlated fluctuations of order parameters at many different points in space.  Their result was simple but powerful.   Consider the result of calculating the average of the product of $m$ local order parameter operators at $m$ different positions, ${\bf r}_m$, in a system at the critical point. (All differences between positions of the operators should be large in comparison to the distance between neighboring sites or the range of forces.) Compare this average with the same correlation function calculated at the positions $\ell \times {\bf r}_m$.  All that has been done is to change the length-scale on which the correlations have been defined. Patashinskii and Pokrovsky then argued that this change in scale was an invariance of the system so that the two correlation functions will have precisely the same structure\cite{PP}, and differ by a factor $\ell^{-m x}$.   A similar rule, with a different index holds, for other kinds of fluctuating quantities near the critical point\cite{PP1}.     These authors succeeded in getting the right general structure of the correlations, but in their initial work\cite{PP} they did not succeed in predicting values of critical indices.  This work pointed the way toward future field theoretic calculations of correlation behavior.  

In parallel, I calculated\cite{LPK} the long-distance form of the spin correlation function for the two-dimensional Ising model by making use of the Onsager solution.  This was the part of a long series of calculations that would give insight into the structure of that model\cite{MW}.  Those insights would be quite crucial in establishing the fundamental theory of behavior at the critical point.

\subsection{Widom scaling}

Benjamin Widom\cite{Widom,WidomII} (see \fig{FW}) reported at the National Bureau of Standards meeting upon a phenomenological theory of the thermodynamics near critical points.  (He studied the liquid-gas transition, but here his results will be stated in the language of the magnetic transition, in which the temperature deviation from criticality is $t$ and the symmetry of the ordering is broken by a magnetic field, $h$.)   If $t$ is zero, the average order parameter, $<\sigma>$,   was experimentally seen  to be proportional to  $h^{1/\delta}$ where $\delta$ is a critical index known to be close to 4.4 in three dimensions\cite{WR} and 15 in two dimensions\cite{67Review}.   As discussed above, if $h=0$ and $t<0$, 
then $<\sigma>$ is proportional to $\pm (-t)^\b$.   
He then said that, near the critical point, no one of these three quantities has a natural size, but instead each one should be measured against the size of the others.  This led him to suggest\cite{Widom}  a general formula for the magnetization near the critical point, that could fit both limiting forms, specifically \be
<\sigma> = h^{1/\delta} g(h^{1/\delta}/t^\b)
\la{Widom}
\ee
where $g$ is a function that would have to be determined experimentally.

   In this way Widom got very concrete   and precise results from his initial requirement that each small quantity, $<\sigma>,h,t$, etc.   be measured against another small quantity.   He was able to predict the index-value to describe how every thermodynamic quantity would go to zero or infinity at criticality.   All these critical indices would then be determined from just the two indices, $\b$ and $\delta$.     

These results were published in a paper\cite{Widom} in the {\em Journal of Chemical Physics.}   In an adjacent paper,  Widom also got a scaling relation\cite{WidomII} for the surface tension, the free energy of the boundary between liquid and vapor, by relating it to the coherence length. To get this, think of  an interface covered by many structures produced by the critical behavior.  One might expect these structures to have a characteristic size of a correlation length, and that each such structure would bring in an extra free energy of order $T$.  Therefore one expects that the entire interface would produce an extra free energy per unit area of order of $T$ times the number of structures that could be placed to fill a unit area, ${\xi}^{-2}$.   In view of \eq{xi}, which ascribes a critical index $\nu$ to $\xi$, we would have a surface tension that varies as $t^{2\nu}$.  This result derived by Widom must have pleased him very much since it shed light on a difficulty that went back to van der Waals.  The theory of the latter gave a critical index for the surface tension of 1.5, while van der Waals' and later experiments gave results in the range 1.22-1.27.  Widom's new theory offered a hope that this old discrepancy between theory and experiment could soon be resolved.   

An  estimation essentially similar to the one Widom used for the surface tension  indicates that the singular term in the bulk free energy has an expected behavior like $t^{3\nu}$.  This result follows from the idea that the free energy in a three-dimensional system would have its singular part determined by excitations with an energy of order $T$.  These excitations would have a  size equal to the correlation length and a density equal to the inverse cube of the correlation length.  Thus,  the surface tension and the free energy density provide a bridge between an understanding of the correlation length and an understanding of thermodynamic properties.     This bridge is not like anything contained in the mean field theories.  In fact these relations, termed hyperscaling relations, are the most characteristic feature of the renormalization theory that will soon arrive on the scene.

\begin{figure}
\begin{multicols}{3}
\includegraphics[height=4.6cm ]{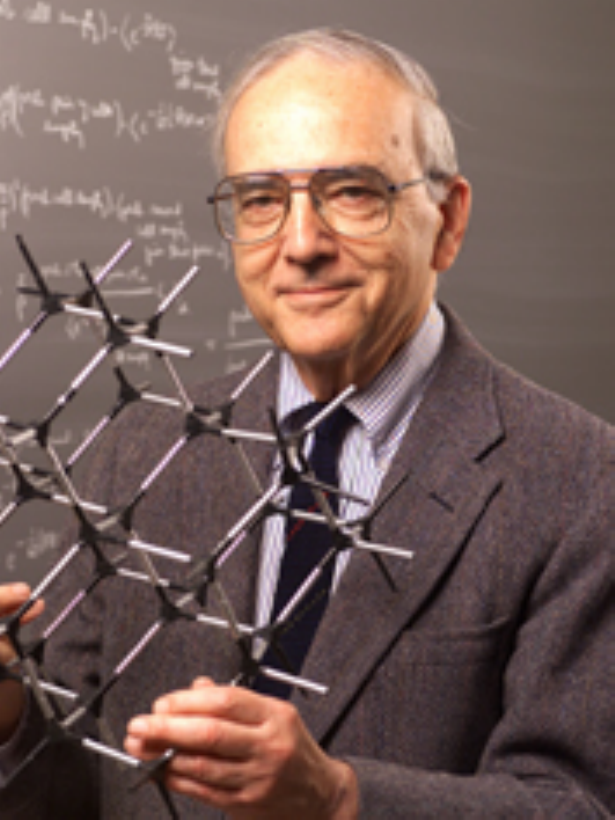}
\includegraphics[height=4.6cm ]{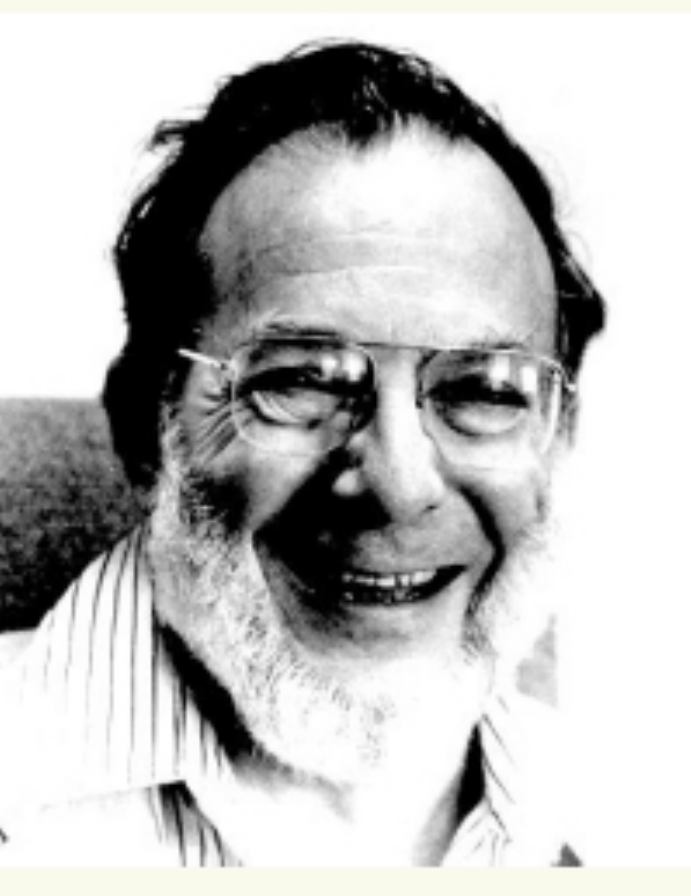}
\caption{Benjamin Widom, left, and Michael Fisher, right. Widom is a Chemistry Professor at Cornell.  Fisher has been at King's College (London), Cornell, and the University of Maryland. }
\la{FW}
\end{multicols}
\end{figure}    

\subsection{Less is the same:  Block transforms and scaling \la{Block}}
Widom supplied much of the answer to questions about the thermodynamics of the critical point.   I then supplied a part of the strategy for deriving the answer\cite{LPK1966}.  I will now describe the method in a bit of detail since the calculation provides some insight into the structure of the solution.  

Imagine calculating the free energy of an Ising model near its critical temperature based upon the interactions incorporated in Ising's Hamiltonian function for the problem.     The result will depend upon the number of lattice sites, the temperature deviation from criticality and the dimensionless magnetic field.   Next imagine redoing the calculation using a new set of variables constructed by splitting the system into cells containing several spins and then using new spin variables,  each intended to summarize the situation in a block containing several old spin variables. (See \fig{block}.) To make that happen one can, for example,   pick the new variables to have the same direction as the sum of the old spin variables in the block and the same magnitude as each of the old variables. The change could then be represented by saying that the distance between nearest neighboring lattice sites would change from its old value, $a$, to a new and larger  value, $a'$.  (See \fig{block}, in which the lattice constant has grown by a factor $\ell=3$.)  In symbols, the change is given by
\bsubs  \la{scaleE}
\be
a' =\ell a
\la{ell}
\ee

  One can then do an approximate calculation and set up a new ``effective''  free energy calculation that will give the same answer as the old calculation based upon an approximate ``effective'' Hamiltonian making use of the new variables.     Near the critical point, one could argue on the basis of universality\footnote{I used \eqs{scaleE}, but I did NOT make an explicit argument based upon universality in my paper in which I first applied this block transformation.  My discussion would have been much stronger had I the wisdom to do so. But wisdom often comes after the fact.} that the new Hamiltonian could be written to have the same structure as the old one.  However, near criticality,  the new parameters in the effective Hamiltonian, the number of lattice sites, the temperature deviation from criticality, and the dimensionless magnetic field all are proportional to the corresponding old parameters. This change can be represented by writing
  \bea
N' &=(\ell)^{-d} N    \la{N} \\
h'&=(\ell)^{y_h} h  \la{h} \\
t' &= (\ell)^{y_t} t \la{t}  
\eea
\la{scalings}
\esubs 
In the first of these statements, \eq{N},  $N$ is the number of lattice sites in the $d$ is the dimension of the lattice. The equation simply describes how the number of sites depends upon the spacing between lattice sites.     

 The other two equations are far, far less simple.   \eq{h} says that the new situation has a symmetry breaking field of the same sign as the previous one.  That would be a reflection of the fact that both situations would have the same kind of ordering.  The coefficient, $(\ell)^{y_h}$, might be derived after some sort of statistical mechanical analysis of the situation.  It is, as it stands, just a number defined by the result of that calculation and one that might depend upon the exact way in which we chose to define the new spin variable.
 
The equation for the new value of the new deviation from criticality, $t=K_c-K$, could be described in similar terms.  It is reasonable to assume that if the original system is at its critical point, so is the new description obtained after the block transformation.  Further it is reasonable to argue that the transformation should engender no singularities, thus requiring that  a new temperature-deviation from criticality would have a linear dependence upon the old deviation.  So the remaining point is to calculate the coefficient in the linear relation, and express it in the special manner given in \eq{t}.

   The proportionality in \eq{t} and \eq{h} are representations of scaling, and the coefficients in the linear relations define the scaling relations among the variables.   
Note that here scaling is viewed as a change in the effective values of the  thermodynamic parameter produced by a change in the length scale at which the system is analyzed.  The length scale must be irrelevant to the determination of the eventual answer and must drop out of the final result for the free energy.   It is this dropping out that gives the empirical relations proposed by Widom.  These scaling relations then give a theory with all the empirical content of Widom's work\cite{Widom}, but backed by the outlines of a conceptual and calculational scheme.

This theoretical work of references \cite{PP,Widom,LPK1966} was well-received. The review paper of \cite{67Review} was particularly aimed at seeing whether the new phenomenology agreed with the experimental data. It reviewed most of the recent experiments but missed large numbers of the older ones that are included in references\cite{Levelt-Sengers,Domb}.   All of this activity validated the consideration of the critical region as an appropriate subject of study and led to a spate of experimental and numerical work, but hardly any further theoretical accomplishments until the work of Wilson\cite{Wilson}.

\begin{figure}
\begin{multicols}{2}
\includegraphics[height=8cm ]{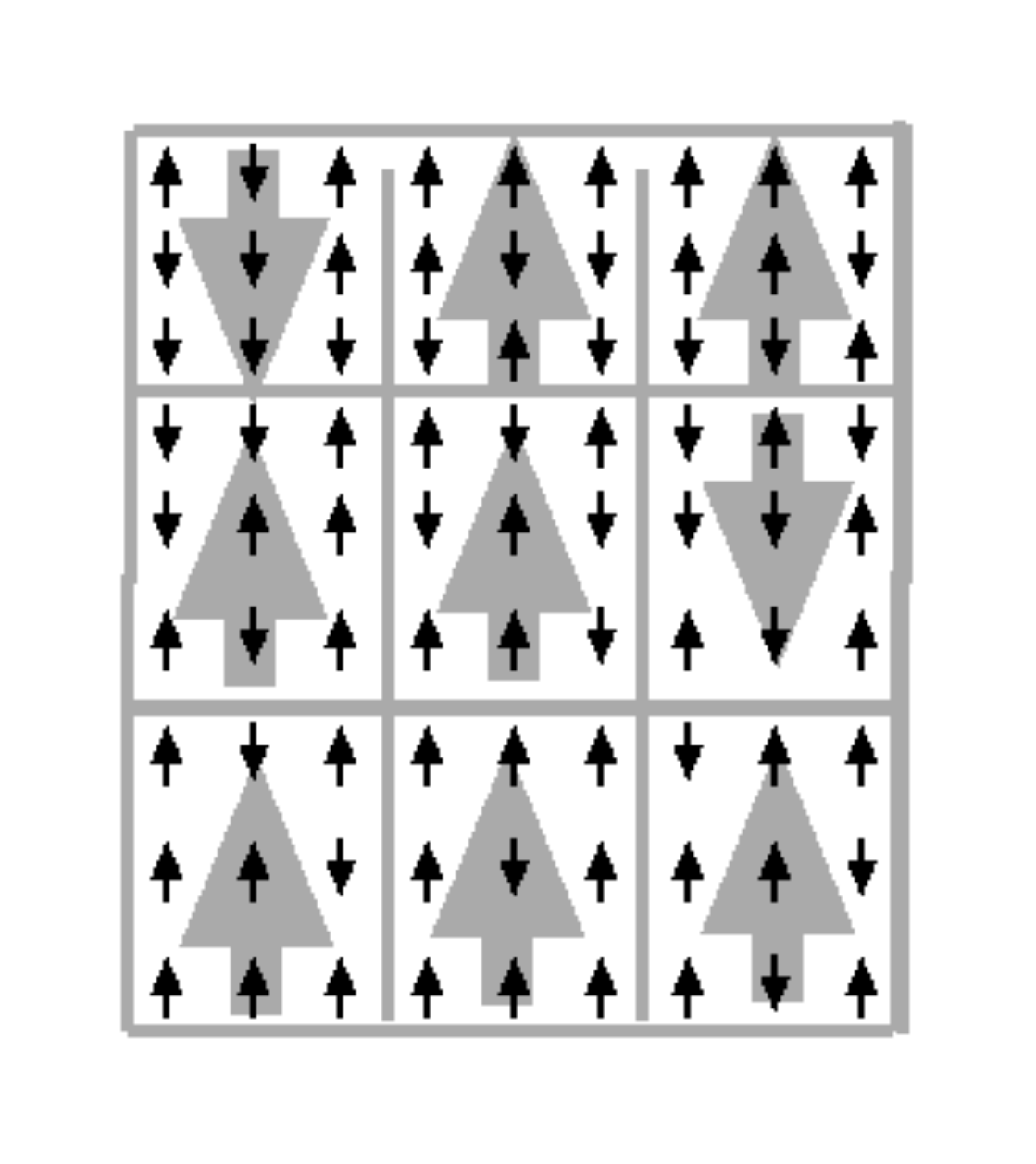}
\caption{Making blocks. In this illustration a two-dimensional Ising model containing 81 spins is broken into blocks, each containing 9 spins.  Each one of those  blocks is assigned a new  spin  with a direction set by the average of the old ones. We imagine the model is reanalyzed in terms of the new spin variables.  }
\la{block}
\end{multicols}
\end{figure}  
\subsection{Physical space; Fourier space}
Before entering into Wilson's construction of the renormalization group theory, I should touch upon a point of technique.

There are two traditional ways of setting up a Hamiltonian or free energy that will then provide a microscopic description of the system.  One way is in coordinate space, the real $XYZ$ space in which you and I live.  This setup is the one we used for the Ising model, the Landau theory,  and for the description of the previous subsection.  It is relatively easy to visualize and the most effective method for problems in low dimensions, specifically for phase transitions in two dimensions.

The other method employs Fourier transforms.  It represents every variable in terms of its Fourier transform. For example the order parameter field of the Landau theory has a transform
\be
{\psi (\k)}= \int~d\r ~ \Psi(\r) e^{-i\k\cdot\r}.
\ee    
The integral covers a space of dimensionality  $d$.   Using $\psi (\k)$ as our basic statistical variable the Landau free energy may be written as
\bea
-F/T &= \int ~ \frac{d{\bf k}}{(2 \pi)^d} 
 \Big[ A h({-\bf k}) {\psi (\k)}
   +B~|{\psi (\k)}|^2  -  D~k^2 |{\psi (\k)}|^2\nonumber  \\
   & +  C \int ~ \frac{d{\bf m}}{(2 \pi)^d} \int ~ \frac{d{\bf n}}{(2 \pi)^d} \int ~    ~{\psi (-\k-{\bf m-n})} {\psi (\bf m  )} {\psi (\bf n  )} {\psi (\bf k  )} 
         \Big]
\la{ftilde}
\eea 
This form in \eq{ftilde}  is used to reach beyond mean field theory and take into account possible fluctuations in the local variables that describe the system.   To do this, one uses $F/T$  as a kind of  a kind of Hamiltonian for phase transition problems.  In this use, the $\k-$ space is divided into small pieces and ${\psi (\k)}$ is taken to be an integration variable in each piece. In this context the expression in \eq{ftilde} is called the Landau-Ginzburg-Wilson free energy.

The $k$-shell integration just described is easily performed if the free energy includes only linear and quadratic terms in the variable, $\psi (\k)$.  The fourth-order term provides a problem, one that can be attacked by using the renormalization method.   The term involving $k^2$ ensures that the contribution to the integral for the highest values of $k$ will be small and relatively easily controlled.  So, one successively  integrates over shells in $\k-$space, starting from the highest values of $|\k|$, and working downward.   As each integral is done, one stops and regroups terms to bring everything back close to the form of the original Landau-Ginzburg-Wilson free energy.  As one does this, the coefficients multiplying the various terms change.

The $\k-$space method is particularly appropriate for higher dimensions, going down to roughly three dimensions.  It is the usual method of choice in particle physics.    In statistical physics, Wilson and Fisher\cite{Wilson-Fisher} have done a very convincing calculation in which they analyze the behavior near four dimensions by assuming that the  fourth-order term is quite small.   (See the discussion of $\e$-expansion in \se{epsilon} below.)  

Both real-space and $\k-$space methods have added considerably to our understanding of phase transitions.  I use the former to describe the concept of renormalization since I find it more natural to think about phenomena  in real space rather than Fourier space.  
 In particle physics, however, our basic conceptualization is based upon, naturally enough, particles.   These are best followed in $\k-$space, since the $\k$  labels the momentum of  particles.  So the two different formulations are complementary, with the best applications to problems in different dimensionalities and indeed to different fields of science. 

The extended singularity theorem, of course, applies equally in both the real-space and the Fourier-space formulations.   In real-space, in order to have the potential for generating singularities, and thereby phase transitions, the system must be infinite in two or more dimensions.  In Fourier-space, the corresponding statement is that two or more components of the ${\bf k}$-vector must extend to infinity.  The remaining requirement in either formulation is that the renormalization must lead to a non-trivial fixed point, one with infinitely large values of some of the couplings.

\subsection{The Wilson revolution\la{Wilson}}
Around 1970, these concepts were extended and combined with previous ideas from particle physics\cite{SP,GML} to produce a complete and beautiful theory of critical point behavior, the renormalization group theory of Kenneth G. Wilson\cite{Wilson}.  (See \fig{Wilson}.) The basic idea of reducing the number of degrees of freedom, described in reference\cite{LPK1966}, was extended and completed.  

\begin{figure}
\begin{multicols}{2}
\includegraphics[height=5cm ]{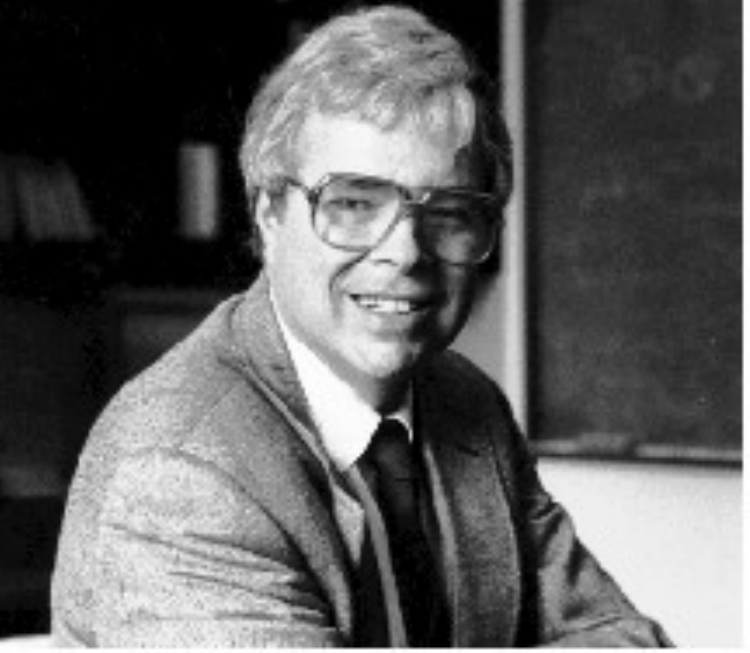}
\caption{Kenneth G. Wilson.     His graduate studies were carried out at the California Institute of Technology where he did a thesis under Murray Gell-Mann, a major contributor to early work on renormalization in particle physics.   This was followed by a Junior Fellowship at Harvard, a year's  stay at CERN, and then an academic appointment at Cornell.  The renormalization group work was done while Wilson was at Cornell.    } 
\la{KWilson}
\end{multicols}
\end{figure}

Wilson, in essence,  converted a phenomenology into a calculational method  by introducing  ideas not present in the earlier phenomenological treatment\cite{LPK1966}:
\begin{itemize} 
\item Instead of using a few numbers, e.g.  $t,h$, to define the parameters multiplying a few coupling terms, he extended the list of possible couplings to include all the kinds of  terms that might be found in the Hamiltonian of the system.  Thereby it became automatically true that the renormalization would maintain the different coupling terms, but only change the size of the parameters which multiplied them.    
\item  Wilson considers indefinitely repeated transformations, as in the earlier particle physics work.   Each transformation increases the size of the length scale.   In concept, then, the transformation would eventually reach out for information about the parts of the system that are infinitely far away.  In this way, the infinite spatial extent of the system became part of the calculation.    The idea that behaviors at the far reaches of the system would determine the thermodynamic singularities were thence included in the calculation. 
\item Furthermore, Wilson added the new idea that a phase transition would occur when the transformations brought the coupling to a {\em fixed point}.   That is, after repeated transformations, the couplings all would settle down to a behavior in which further renormalization transformation would leave them unchanged.   
\item 
 Finally, at the fixed point,  the correlation length would be required to be unchanged by renormalization transformations.  The transformation multiplies the length scale  by a factor that depends upon the details of the transformation.  Wilson noted that there are two ways that the correlation length might be unchanged.  For transformations related to a continuous transition, the correlation length is infinite, thence reflecting the infinite-range correlation.  For transformations related to first-order transitions, the correlation length is zero,  reflecting the local interactions driving the transition. 
\end{itemize} 

A very important corollary to the use of repeated transforms is the idea of {\em running coupling constants}.   As the length scale changes, so do the values of the different parameters describing the system.   In the earlier field theoretical work\cite{SP,GML}, the important parameters were the charge, masses, and couplings of the ``elementary'' particles described by the theory.  These parameters were specified at the beginning and were, in no sense, the outcome of the renormalization calculation.   The change in length scale then changed these prespecified parameters from the ``bare'' values appearing in the basic Hamiltonian  to renormalized values that might be observed by experiments examining a larger scale.   In the new theory,  the number and type of relevant parameters were determined by the outcome of the renormalization calculation.   
 
The use of renormalized or ``effective'' couplings was current not only in particle theory, but also in quasiparticle theories that are pervasive in condensed matter physics\cite{AndersonConcepts}.    In these theories one deals with particles that interact strongly with one another.  Nonetheless, one treats them using the same Hamiltonian formalism that one would use for non-interacting particles.   The only difference from free particles is that the Hamiltonian is allowed to have a position and momentum dependence that reflects the changes produced by the interactions.   Here too, as in the early particle physics work, the quantities to be renormalized are prespecified.  In contrast, see \se{relevant}, Wilson's renormalization calculation determines what is to be renormalized. 

\subsubsection{Different kinds of fixed points}
Wilson's theory gives three different kinds of fixed point corresponding to three qualitatively different points in phase diagrams. For the {\em weak coupling fixed point}, couplings can go to zero and the correlation length goes to zero. The symmetry represented by the order parameter will remain unbroken.  This kind of fixed point describes all areas of the phase diagram that do not touch a phase transition.  For the {\em strong coupling fixed point}, some couplings will go to infinity  and the correlation length goes to zero.   Here the basic symmetry represented by the  order parameter gets broken by at least one non-zero coupling that violates that symmetry.  This kind of fixed point describes all areas of the phase diagram that  touch a first-order phase transition.  For the {\em critical fixed point} couplings remain finite, the symmetry remains unbroken, and a correlation length goes to infinity.  

\subsubsection{Different scalings: relevant, irrelevant, marginal\la{relevant}}
Since the Wilsonian point of view generated the renormalization of many different couplings, it became important to keep track of the different ways in which the couplings in the free energy would change as the length scale changes.  This work starts with an eigenvalue analysis.  One takes linear combinations of couplings arranging the combinations so that, after a renormalization, every combination reproduces itself except for a multiplicative factor.   In other words, these linear combination make every linear combination of couplings obey an equation like the ones in \eqs{scalings}, so that every combination, $s$, obeys 
\be
s'=\ell^{y_s} s
\la{ydef} 
\ee    
The different combinations are then classified according to the values of the index, $y_s$.  Many of the combinations of couplings have an index that is complex, but I shall focus upon the cases in which the index is real.  That is the most significant situation, anyhow.  There are then three possibilities
\begin{itemize}
\item   Relevant,  $y_s$ greater than zero.   These are the couplings like $t$ and $h$ that grow larger as the length scale is increased.   Each of these will, as they grow, push one away from the critical point.  In order to reach the critical point, one must adjust the initial Hamiltonian so that these quantities are zero.
\item Irrelevant, $y_s$ less than zero.   These coupling will get smaller and smaller as the length scale is increased so that, as one reaches the largest length scales, they will have effectively disappeared
\item Marginal, $y_s$ equal to zero.   
\end{itemize} 
The last case is rare.  Let us put it aside for a moment and argue as if only the first two existed.

\subsubsection{Universality classes\la{Universality} }

To study critical phenomena based upon renormalization transformations, one sets all the relevant combinations of couplings to zero and then does a sufficient number of successive renormalizations so that all the irrelevant combinations have also disappeared.  We thus end up with a unique {\em fixed point} independent of the value of all of the irrelevant couplings.   The act of renormalization is a sort of focusing in which many different irrelevant couplings fade away and we end up at a  a single fixed point representing a whole multi-dimensional continuum of different possible Hamiltonians. These Hamiltonians form what is called a {\em universality class.}
Each Hamiltonian in its class has exactly the same critical point behavior, with not only the  same critical indices, but also the same long-ranged correlation functions, and the same singular part of the free energy function.  

The identity among different problems is not just a theoretical artifact.  As first pointed out by Lee and Yang\cite{Lee-Yang,Lee-YangII}, the Ising model, single axis ferromagnets, and the liquid-gas phase transition all show identical critical properties.   The theory makes these critical properties vary with dimension, and experiments bear out the predicted universality in the two observable cases: $d=2$ and $d=3.$ As another example,    XY ferromagnets have a two component order parameter, with the same symmetry properties as superfluids, with their complex order parameter. In three dimensions, the critical behavior of XY ferromagnets and superfluids fall into the same universality class, but, of course,  one that is quite different from the universality class that applies equally to the critical behavior of the liquid-gas phase transition and the Ising model.   

    This universality-class idea has been applied to many different problems beyond critical phenomena\footnote{I must admit to a certain pride connected with universality.  The 1967 review paper\cite{67Review} in which I participated was organized about universality classes. One might also notice that word ``universality'' was translated from the Russian and was borrowed by me from the conversation of Sasha Polyakov and Sasha Migdal and thus imported into the English language, with this specific meaning\cite{me}. Alternatively,  one might argue that universality was a product of many different authors, including Robert Griffiths\cite{Griffiths} as well as the entire King's college school\cite{Domb}.  }.  Whenever two systems show an unexpected or deeply rooted identity of behavior they are said to be in the same universality class. 

There are, of course, many different universality classes corresponding to different dimensionalities, different symmetries of the order parameter, and to different stability properties of the fixed points.

Before leaving this subject, focus once more on the possibility of a marginal behavior.   In the marginal case, we have a coupling that does not vary under renormalization.   That kind of coupling can produce critical properties that vary continuously as some parameter is varied.  For example a pair of coupled Ising models living in the same space show a marginal behavior of this kind\cite{Kadanoff-Wegner}.  A slightly different marginal behavior is shown by the XY model in two dimensions\cite{KT,ZH}.  
        
\subsection{New concepts \la{concepts}}
The renormalization revolution and the phenomenological work that preceded it included several very important new ideas that were applied not only to critical phenomena but also to many other situations.  These new concepts include, in addition to the analysis of couplings described above, 
\begin{itemize}
\item Fluctuations.  The behavior at the critical point is determined by fluctuations in all kinds of physical quantities, with the order parameter being the most important.
\item Scaling.  Near criticality, fluctuations occur over a very wide range of length scales.  The near-critical couplings like $t$ and $h$ can each be measured against powers of the relevant length scale.   The renormalization group calculation gives the values of critical indices to describe the scaling behavior of the physical quantities that define the critical point. 
\item Universality classes.  The universality idea was used to classify all the different possible behaviors at critical points.  There are only a limited number of different universality classes.  When different phase transition fall into the same class, all aspects of their critical behavior then turn out to be identical. For example, the liquid-gas phase transition and the Ising model are in the same universality class.   
\item Symmetry.    Different symmetries and different numbers of dimensions imply different universality classes.  Thus a superfluid transition on a surface is quite different from one in bulk.                  
\end{itemize}   

  In one sense the renormalization group is rather different from anything that had come before in statistical physics, and by extension in other parts of physics as well.  Previous work in statistical physics  had emphasized finding the properties of {\em problems} defined by statistical sums, each sum being based upon a probability distribution defined by particular values of coupling constants like $K$ and $h$.  Such sums would be called {\em solutions} to the problems in question. In the renormalization group work the emphasis is on connecting problems by saying that different problems could have identical solutions.  The method involved finding different values of couplings  that would then give identical free energies and other properties. These set of couplings would then form a representation of a universality class.    All the interactions that flow into a given fixed point in the course of an infinite number of renormalizations belong to the  universality class of that fixed point  
  
A universality class would give a solution, in the old sense, if one finds within the class a set of couplings so simple that the solution is obvious. This is what happens when the running couplings produce  infinitely weak interactions, thereby producing  a weak coupling fixed point.  A strong coupling fixed point might also be trivial if no important symmetry remains after the order parameter takes on a non-zero value.  However, a first-order phase transition might produce a nontrivial situation with quite a bit of remaining symmetry.  In that case, further analysis is necessary before one can get anything like a solution in the old sense of the word. 

Finally,  a critical fixed point  is not really a ``solution'' in the old sense.  It gives us values of critical indices and describes scaling behavior,  which can then be used to infer many of the qualitative properties of a solution.  But many of the details of the old-sense solution may not be available from a knowledge of the fixed point alone.

\subsubsection{Extended singularities revisited}
The renormalization group has an entirely different spatial structure from that of mean field theory. The difference can best be seen by comparing the Ising model mean field theory of \se{Mean} with the block spin formulation of \se{Block}.  

In the mean field formulation, the value of an average magnetization at point  ${\bf r}$ is determined, first of all, by the values of the magnetization at points connected by bonds to the initial point.  These are then determined, in turn, by magnetizations at points connected these new points by bonds. This bonding process might continue indefinitely or terminate after a finite number of steps.  Mean field theory does not have the right spatial structure for the correct prediction of phase transitions.

In contrast, the blocking procedure of the renormalization group determines the couplings in a given region,  in the first analysis, by the effects of couplings in a region of size $\ell$ larger.  The blocking then reaches out in geometric progression to regions each expanded by a factor of $\ell$. 

Of course, the block transformation reach out more quickly and effectively than do the steps of the mean field calculation.  But that is not the main difference.  The mean field theory can have a pseudo-phase transition determined by just a few couplings.   On the other hand, if the block transformation ever reaches out and sees no more couplings in the usual approximation schemes(\cite{NL}) that will signal the system that a weak coupling situation has been encountered and will cascade back to produce a weak coupling phase.  Hence the blocking approach has the potential of using the right fact about the spatial topology to determine the possibility of a phase transition. 

By this argument the extended singularity theorem suggests that phase transitions are triggered by a very elegant mathematical juxtaposition put before us by mother nature.   On one hand, the phase transition is connected with a symmetry operation built into the microscopic couplings of the system.  For example,  the ferromagnetic based upon the breaking of a symmetry in the possible direction of spins.  One the other hand the phase transitions also make use of the extended topology of an system that extends over an effectively infinite region of space.   This coupling of microscopic with macroscopic has an unexpected and quite breathtaking beauty.

\subsubsection{Flows and flow diagrams}
A renormalization operation differs in a very deep sense from all the usual calculations performed within the statistical mechanics of Boltzmann and Gibbs. In statistical mechanics you start with a  statistical ensemble, usually defined with a Hamiltonian, and use that ensemble to calculate an average.  In a renormalization operation, you start with a statistical ensemble, usually defined by a Hamiltonian, and you calculate another ensemble, often described by a Hamiltonian containing different couplings.  In one case the calculation is, in brief, {\em ensemble generates averages}, in the other,  the calculation is {\em  ensemble generates ensemble.} This is quite a substantial difference.

The part of mathematics that goes with standard statistical mechanics is probability theory.  The part of mathematics that goes with renormalization is called ``dynamical systems theory,'' and describes how things change under transformations.  The concepts of a fixed point and of a basin of attraction belong to dynamical systems theory rather than probability theory.

Dynamical systems theory is often used to describe continuous changes, as for example the changes in a mechanical system as its state changes  in time.  For the purposes of this section, I will speak as if all renormalization transformations were continuous changes produced by an infinatesmal increase in a basic length.  Thus, the transformation will be, $a \rightarrow a +d\ell$. Then, every coupling also undergoes an infinitesimal change $K \rightarrow K +dK $.  In particle physics this kind of approach has the name of the Callen-Symanzik equation\cite{Callen,Symanzik,Symanzik2}.

The simplest kinds of flow pictures look at a single coupling constant, $K$, and how that coupling changes under renormalizations.   In the one-dimensional Ising model, depicted in  \fig{IsingFlow1d}, each renormalization makes the coupling weaker.  Thus the coupling flows toward the weak coupling fixed point at $K=0$. In contrast, the flow in \fig{IsingFlow2d}  describes the two-dimensional Ising model.  The flow is zero at the critical fixed point.  To the left of the critic fixed point all couplings flow toward the weak coupling fixed point at $K=0$; to the right all flows go toward the strong coupling point at $K= \infty$.     This diagram describes a system with a single critical point, but a total of three fixed points.    
\begin{figure}
\includegraphics[height=3cm ]{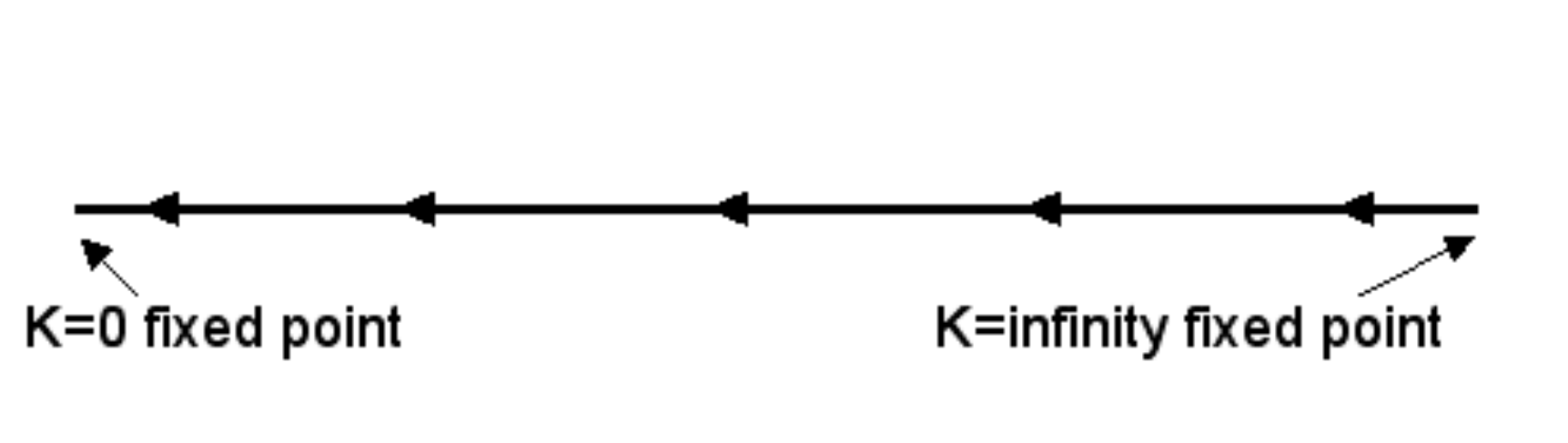}
\caption{Flow diagram for one-dimensional Ising model.  Renormalizations weakens the coupling and pushes it toward a weak coupling fixed point at $K=0$.   }
\la{IsingFlow1d}
\end{figure}

\begin{figure}
\includegraphics[height=3cm ]{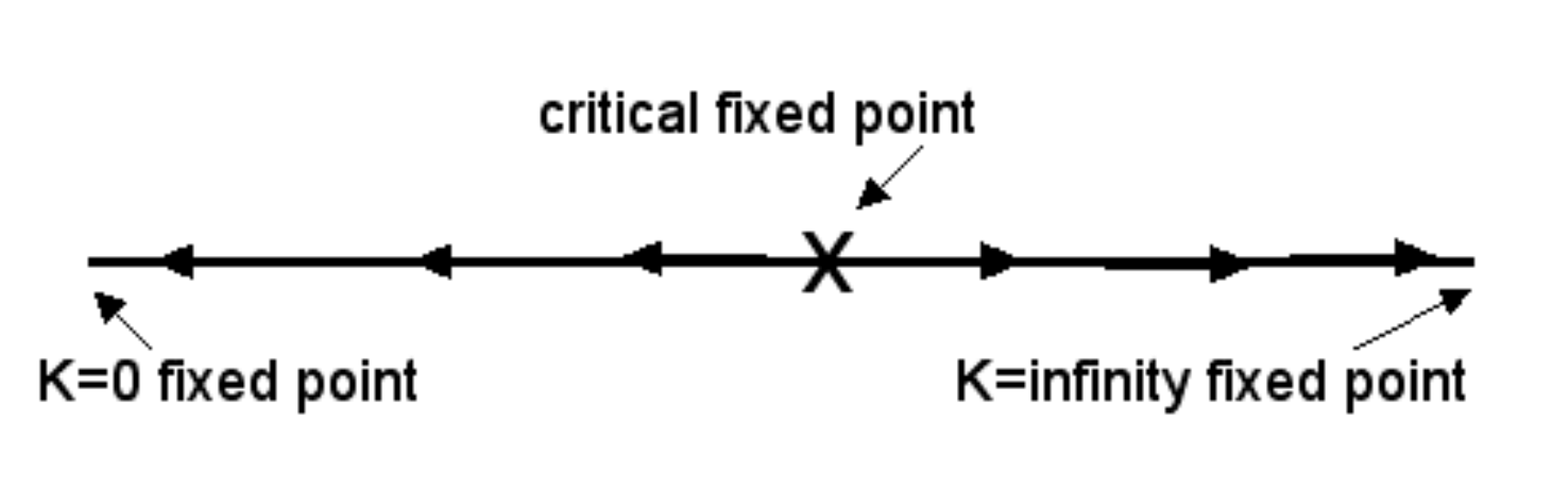}
\caption{Flow diagram for two-dimensional Ising models.  There is a criticial fixed point at $K=K_c$.  For initial couplings weaker than this value,  renormalizations further weaken the coupling and push it toward a weak coupling fixed point at $K=0$.  Conversely, if the initial couplings are stronger than $K_c$, renormalizations produce a flow toward a strong coupling fixed point, describing a ferromagnetic state.  }
\la{IsingFlow2d}
\end{figure}

We have come a long way from the starting point set by Boltzmann and Gibbs.  Solutions to problems in statistical mechanics has here been described
in terms of renormalization group flows, universality classes, and types of fixed points.  This new language has become important in statistical physics and has been extended to applications well beyond the situations described here.  This language, derived from statistical mechanics, has become even more pervasive in particle physics, where coupling constants run everywhere. A calculational method is more than a way of putting symbols on paper.   It provides a way of looking at, and conceptualizing, nature.


\subsubsection{The renormalization group is not a group  }
Although the renormalization operation is usually described as a part of a group, it  actually forms a semigroup.  A group is a set of operations  with three  characteristics:
\begin{itemize}
\item  Two operations in the group, taken in succession, produce another group operation.
\item The group contains an element called the identity, which has the effect of changing nothing whatsoever.
\item  For each operation in the group there is, as part of the group,  an inverse operation,  so that when you successively perform  the operation and its inverse, that pair of operations produces the identity element.   
\end{itemize}
A semigroup lacks the third characteristic.    Once you have performed a group operation you cannot necessarily undo that operation.

The reason that renormalization produces a semigroup is that a block transformation (see \se{Block}) loses information. After the transformation, the system contains fewer lattice sites and so can hold less ``information.''  Some irrelevant couplings, which could be seen before the transformation have simply disappeared. (These couplings have the index value $y=-\infty$.  In addition, there are other kinds of couplings, called ``redundant''  that do not affect the free energy and so disappear without a trace in the course of a renormalization. )  Both kinds of coupling make renormalization a semigroup operation.

This characteristic is important because it eliminates or greatly reduces  the possibility of finding the small-scale Hamiltonian of the system by looking at large-scale phenomena.  Before the use of renormalization methods scientists often though that a sufficiently accurate and detailed study of a system, albeit a study conducted on a large length scale, could determine all the basic laws governing the system, down to the smallest scale.    In practice the disentanglement of microscopic laws always proved to be hard.  But, in principle, it was always assumed to be possible.    But the renormalization group theory says that information will disappear in the process of changing length-scales. Even ordinary irrelevant operators have effects that disappear with exponential rapidity in the course of a renormalization transformation.  These do not produce, in principle, a disappearance of information but they make it well-nigh impossible to reconstruct a small-scale Hamiltonian from large-scale data.  

Thus, it is often wise for a scientist to consider the macroscopic world to either be interesting  on its own merits, or not at all.  We cannot expect to gain very much reliable information about the microscopic world from the direct examination of the macroscopic one. (We can, on the other hand, gain lots of insights that can be employed to help construct theories about microscopic phenomena.)   Happily, the existence of quite non-trivial fixed point behavior shows that the macroscopic world has quite enough interest to engage the imagination of an intelligent scientist.  

The previous paragraph was constructed to describe a situation in which there is but one macroscopic world to understand.  In fact, there are many.  Each fixed point described by the renormalization group defines its own unique little world.  Thus a plasma is different from a magnet that in turn is different from a set of defects in a crystal lattice.   Each of these physical systems defines its own little world with rules and physical laws.  This point will be developed further in the next section.  

\subsection{A calculational method defines many worlds}
Wilson, in essence,  converted a slightly vague phenomenological theory into a well-defdined calculational method.  So are they all:  classical mechanics, quantum mechanics, statistical mechanics, field theory,...., all calculational methods. But they are also each complete descriptions of some ``sub-universe.'' As such, they each engender their own world and their own philosophy. As you can see from the set of ideas outlined in \se{concepts}, the renormalization group built its own set of philosophical perspectives, which then displayed both condensed matter and particle physics in a new and different light. 

However, there is an additional deep sense in which the renormalization group defines its own worlds.   Each fixed point is a microcosm, worthy of study on its own merits.     I spent many years studying the two-dimensional Ising model near criticality, learning particularly about the behavior of its correlation functions.  I feel the richer for the experience.  

How can I justify viewing each renormalization group fixed point as its own little world?   Each fixed point has its own basin of attraction defined by its very large set of irrelevant couplings.  This basin of attraction is the region in the space of possible Hamiltonians that will eventually produce flows into our particular fixed point. We can visualize the flow as the simultaneous diminution of each of these variables.  When two or more of these variables interfere, we can see non-linear effects in the flow toward the fixed point.  Since we expect that our listing of irrelevant variables is a complete list, we expect that two variables acting together will produce an effect that we can describe as a summed effect of the variables on our previous list.  In this way, we get a kind of multiplication table in which the product of any two variables is a sum of the others with specified coefficients. Such a multiplication table is what the mathematicians call an algebra.  This algebra defines in the deepest way what is happening in the phase transition. 

The algebras that have actually been studied are a little deeper than the one just described. They are produced not just by the couplings, but by the specifications of the couplings in local regions of the system.   Therefore the algebras combines the properties of space with the properties of the particular fixed point.   They have been most richly studied in two dimensions in which the spatial part common to all these algebras is called the Virasoro algebra\cite{Virasoro} and the application to critical phenomena was championed by Dan Friedan, Dongan Qu, and Stephen Shenker\cite{FQS}.  This approach also plays an important role in string theory.   

Each fixed point has its own unique algebra that describes the structure of the local correlations determining the fixed point behavior.  This algebraic structure was originally emphasized by Kenneth Wilson\cite{OPE} and myself\cite{OperatorAlgebra} under the names of short distance expansion and operator product expansion.

\subsection{Building upon the revolution\la{Building} }
This Wilson renormalization theory  provided a basis for the development of new methods that could be used for building an understanding of critical phenomena. It provided a framework into which one could fit a variety of different theories and physical problems. 
\subsubsection{The $\e$-expansion\la{epsilon}}

One example is the {\em $\epsilon $}-   expansion of  Wilson and Fisher\cite{Wilson-Fisher}. Here $\e$ means dimension minus four.  This calculational method focuses upon the dependence of physical quantities upon dimension.  It uses renormalization transforms near four dimensions, where mean field theory is almost, but not quite, correct\footnote{The idea of variable dimensionality is also used in particle physics under the name dimensional regularization.  One of its earliest applications was in the  work of 't Hoooft and Veltman\cite{tH,HVa,HVb} that proved that the gauge theory of strong interactions was renormalizable.}.  The idea of using the dimension of the system as a continuously variable parameter seems a bit strange at first sight. However, in the momentum-space representation of statistical ensembles, each term in a perturbation expansion can be evaluated for all integer values of the dimension and then the analysis can be continued to all values of the dimension, including non-integer values.   

When applied near four dimensions this method allows an almost exact analysis of the fixed point behavior.  Near the fixed point,   the non-trivial terms in the free energy, like the term proportional to $D$ in \eq{ftilde}, go to zero as the dimension approaches four.       Because of this simplification, the method gives  quite accurate results for  critical behavior near four dimensions.  Further, it provides a series expansion that gives useful answers for many different models in three dimensions.  The close correspondence of theory and experiment helped to convince people that both the variable-dimension method and the renormalization method were valid.   The way had been opened for an explosion of new calculations and new understandings.

\section*{Acknowledgment} Some of the material in this review was prepared and delivered in a talk an an APS 2002 meeting commemorating the hundredth anniversary of the death of J. Willard Gibbs.  Additional material  was first prepared for the Royal Netherlands Academy of Arts and Sciences in 2006.  Still more material in this paper appeared in part in a talk at the 2009 Seven Pines meeting on the Philosophy of Physics under the title ``More is the Same,   
Less is the Same, too;  Mean Field Theories and Renormalization.''  These talks have appeared on the authors' web site\cite{LPKwebsite} since then.   This Seven Pines meeting was generously sponsored by Lee Gohlike. The paper also incorporates material from ``More is the Same'' published in J. Stat. Phys. in 2009\cite{I}. 

This work was was supported in part by the University of Chicago MRSEC program under NSF grant number DMR0213745.  It was completed during  visits to the Perimeter Institute,  which is supported by the Government of Canada through Industry Canada and by the Province of Ontario through the Ministry of Research and Innovation, and by the present NSF DMR-MRSEC  grant number 0820054.

I had useful discussions related to this paper with  Tom Witten,  E. G. D. Cohen, Gloria Lubkin,  Ilya Gruzberg, Wendy Zhang, Roy Glauber, Yitzhak Rabin, Sidney Nagel, and Subir Sachdev.  I owe particular thanks to Michael Fisher who, as he has done many times, helped me understand this interesting subject.

\bibliographystyle{plain}
\bibliography{philo2}
\end{document}